\documentclass[conference]{IEEEtran}
\IEEEoverridecommandlockouts
\usepackage{cite}
\usepackage{amsmath,amssymb,amsfonts}
\usepackage{algorithm}
\usepackage{algorithmicx}
\usepackage{algpseudocode}
\usepackage{setspace}
\usepackage{graphicx}
\usepackage{textcomp}
\usepackage{xcolor}
\usepackage{diagbox}
\usepackage{multirow}
\usepackage{subfigure}
\PassOptionsToPackage{hyphens}{url}
\usepackage{hyperref}
\hypersetup{colorlinks, citecolor=green, filecolor=pink, linkcolor=red, urlcolor=blue }

\makeatletter
\newenvironment{breakablealgorithm}
  {
   \begin{center}
     \refstepcounter{algorithm}
     \hrule height.8pt depth0pt \kern2pt
     \renewcommand{\caption}[2][\relax]{
       {\raggedright\textbf{\ALG@name~\thealgorithm} ##2\par}%
       \ifx\relax##1\relax 
         \addcontentsline{loa}{algorithm}{\protect\numberline{\thealgorithm}##2}%
       \else 
         \addcontentsline{loa}{algorithm}{\protect\numberline{\thealgorithm}##1}%
       \fi
       \kern2pt\hrule\kern2pt
     }
  }{
     \kern2pt\hrule\relax
   \end{center}
  }
\makeatother

\begin{document}

\date{}

\def\BibTeX{{\rm B\kern-.05em{\sc i\kern-.025em b}\kern-.08em
T\kern-.1667em\lower.7ex\hbox{E}\kern-.125emX}}
\title{\Large \bf AxeChain: A Secure and Decentralized blockchain \\ for solving Easily-Verifiable problems}

\author{\IEEEauthorblockN{Weilin Zheng\IEEEauthorrefmark{1},
Xu Chen\IEEEauthorrefmark{1}, 
Zibin Zheng\IEEEauthorrefmark{1},
Xiapu Luo\IEEEauthorrefmark{2},
Jiahui Cui\IEEEauthorrefmark{1}}
\IEEEauthorblockA{\IEEEauthorrefmark{1}School of Data and Computer Science, Sun Yat-sen University, Guangzhou, China}
\IEEEauthorblockA{\IEEEauthorrefmark{2}Department of Computing, Hong Kong Polytechnic University, Hong Kong}

Emails: \{zhengwlin, chenx397\}@mail2.sysu.edu.cn, zhzibin@mail.sysu.edu.cn, hndai@ieee.org, cuijh6@mail3.sysu.edu.cn
}

\maketitle

\begin{abstract}
While Proof-of-Work (PoW) is the most widely used consensus mechanism for blockchain, it received harsh criticism due to its massive waste of energy for meaningless hash calculation. Some studies have introduced Proof-of-Stake to address this issue. However, such protocols widen the gap between rich and poor and in the worst case lead to an oligopoly, where the rich control the entire network. Other studies have attempted to translate the energy consumption of PoW into useful work, but they have many limitations, such as narrow application scope, serious security issues and impractical incentive model.

In this paper, we introduce AxeChain, which can use the computing power of blockchain to solve practical problems raised by users without greatly compromising decentralization or security. AxeChain achieves this by coupling hard problem solving with PoW mining. We model the security of AxeChain and derive a balance curve between power utilization and system security. That is, under the reasonable assumption that the attack power does not exceed 1/3 of the total power, 1/2 of total power can be safely used to solve practical problems. We also design a novel incentive model based on the amount of work involved in problem solving, balancing the interests of both the users and miners. Moreover, our experimental results show that AxeChain provides strong security guarantees, no matter what kind of problem is submitted.
\end{abstract}


\section{Introduction}

Since the debut of Bitcoin\cite{nakamoto2008bitcoin}, blockchain has attracted lots of attentions and been used in various areas. Blockchain systems leverage consensus mechanisms to reach consensus on each block, which contains a set of transactions. Although many consensus algorihtms have been proposed\cite{natoli2019deconstructing}, PoW is the most widely used one. For example, Bitcoin\cite{nakamoto2008bitcoin} and Ethereum\cite{wood2014ethereum}, the top two cryptocurrencies, are using PoW. Unfortunately, PoW has received harsh criticism due to its massive waste of energy for meaningless hash calculation\cite{zheng2018blockchain}. For example, the peak power consumption of Bitcoin network reaches 60 billion kWh/year\cite{bitcoinenergyconsumption}, while Denmark's electricity consumption in 2014 was only 32 billion kWh\cite{electricityconsumption}.

For the energy waste in PoW, the current alternative is Proof-of-Stake (PoS)\cite{king2012ppcoin}. Under PoS, miners compete for the book-keeping right based on their mortgaged stake rather than computing power, and these stake will be confiscated if they are evil. PoS is based on the assumption that the more stake a miner has, the less likely it is to attack the blockchain network. First of all, whether this assumption is reasonable or not remains to be explored. For example, on September 26, 2018, the cryptocurrency exchange Huobi was caught in a scandal of voting control in the EOS\cite{io2017eos} consensus node election\cite{huobiDENIES}. Secondly, the issue of the disparity between the rich and poor caused by PoS is affirmative. At present, there are still many disputes about PoS\cite{poelstra2014distributed, houy2014will}, and its safety and robustness are not superior to PoW. 

Although some other studies translate the energy consumption of PoW into useful work, they have many limitations, such as narrow application scope, serious security issues and impractical incentive model. For example, some studies\cite{king2013primecoin, Gapcoin} replace hash calculation with search of prime sequence. However, their application scope is narrow. While others\cite{oliver2017proposal, chatterjee2019hybrid, amar2019incentive} translate the energy consumption of PoW into solving practical problems like NP problems, we find serious security issues in them. More precisely, under the consensus mechanisms in \cite{oliver2017proposal, chatterjee2019hybrid}, a miner can generate a block and get reward if it solves the NP-complete problem \emph{or} the traditional PoW puzzle. We name such strategy as \textit{binary block-generation strategy (BBGS)}, because it has two ways to generate new blocks. We find that attackers can exploit such mechanisms to make the blockchain system have extremely low availability and security because the generation of new block is seriously affected by the difficulty of problems submitted by users (Please refer to the details in Section \ref{BBGSSecurity}). Apart from the security issues, existing studies lack a reasonable incentive model to protect the fairness of interests between miners and users. For example, in \cite{oliver2017proposal, chatterjee2019hybrid, amar2019incentive}, the reward for solving problem is subjectively determined by users. Since it is difficult to accurately assess the difficulty of the problem, users may worry about paying too much, while miners worry about too little income.

In this paper, we propose AxeChain, which integrates practical problem solving into PoW mining in a secure manner. In particualr, we design the \textbf{Axe blockchain architecture} and \textbf{Virtual machine stack sampling proof algorithm} to address the security issues of \textit{BBGS}. Moreover, we propose a novel incentive model, that assign reward according to the amount of work exactly involved in calculating the problem, to avoid the loss of both the users and miners. It is worth noting that this paper also introduces a more flexible and versatile method for representing general problems in a smart contract, which can expand the application scope of AxeChain. The security analysis of AxeChain in this paper shows that under the reasonable assumption that attackers' power does not exceed 1/3 of the total computing power, the system can obtain 1/2 of the power utilization radio. That is, half of the total computing power can be used to solve the practical problems.

This paper makes the following contributions:
\begin{itemize}
  \item To our best knowledge, AxeChain is the first Strongly Secure and Fully Decentralized blockchain for solving Easily-Verifiable problems. AxeChain is not only a secure decentralized ledger system, but also a decentralized computing platform with strong computing power.
    \item We design a novel incentive model based on the amount of work exactly involved in calculating the problem, which can balance the interests between users and miners.
  \item We design a flexible approach for representing general problems, which can expand the application scope of AxeChain.
\end{itemize}

The rest of the paper is organized as follows. Section \ref{Background} introduces some background and preliminaries of this paper, while Section \ref{BBGSSecurity} analyzes the security issues of BBGS in the existing studies. Section \ref{SystemDesign} and \ref{samplingproof} will introduce two highlights of this paper, which address the above security issues. Then, we introduce a novel incentive model in Section \ref{incentivemodel} and present some experimental results in Section \ref{ExperimentalResults}. Finally,we present some related works in Section \ref{relatedwork} and make a conclusion in Section \ref{conclusion}. 

\section{Background}\label{Background}
\subsection{PoW \& PoS}
PoW in main cryptocurrencies, including Bitcoin\cite{nakamoto2008bitcoin} and Ethereum\cite{wood2014ethereum}, require a large amount of computing resources to reach consensus on each block. In a PoW-based blockchain system, the generation and verification of a block involves a mathematical problem (also called a PoW puzzle). The difficulty in finding the answer to the puzzle is moderate, but verifying the correctness of the answer is easy. Miners must invest a lot of computing resources to find the answer to the PoW puzzle in each consensus round, so the system should cooperate with a incentive model to reward miners. An inappropriate incentive model will cause the loss of computing power and lead to inadequate computing power in the whole network. In this case, it is easier for a malicious miner to obtain more than half of the computing power of the entire network to attack the blockchain, that is \textit{majority-attack}\cite{MajorityAttack}. Currently, there are mainly two different POW consensus mechanisms. 

Bitcoin's PoW is a computationally-intensive consensus mechanism used by Bitcoin. Under Bitcoin's PoW, when mining, miners constantly perform two rounds of ${SHA2(Nonce,block\_header)}$\cite{SHA2} to get a hash (two-round hash is designed to avoid the potential Malleability-Attack\cite{duong2009flickr} of SHA-2). For convenience, in this paper, we call this hash as a workload or a final workload, which used for comparison with the target to determine whether to generate a new block. Only if the workload is less than the target can miners generate a new block. The unidirectionality of the hash algorithm ensures that miners cannot reverse the legal Nonce from the target. Miners can only find a legal Nonce by constantly traversing the Nonce field and doing hash calculation.

Ethash \cite{ethash} is another PoW consensus mechanism used by Ethereum. Unlike Bitcoin's PoW, Ethash is IO-intensive and is designed to withstand Application Specific Integrated Circuit (ASIC)\cite{smith1997application}. ASIC mining rigs makes each hash operation faster and consumes less energy, but it also raises the uptake barriers for peers to participate in consensus. Ethash was originally designed to allow more people to use common computer to participate in consensus without buying ASIC mining rigs. Under Ethash, miners will generate a GB-level directed acyclic graph based on a seed before consensus. When mining, miners perform one round of ${SHA3(Nonce,block\_header)}$\cite{SHA3} to get an initial mixed result ${mix_0}$. Then, based on the indexs generated by ${mix_0}$, miners take fragments from the directed acyclic graph, and then use FNV\cite{fowler2011fnv} to fully mix the fragments with ${mix_0}$ to obtain a new mixed result ${mix_{new}}$. Finally, miners perform another round of ${SHA3(mix_0,mix_{new})}$ to generate a final workload, which is used to compare with the target. Therefore, the bottleneck in Ethash workload generation is to read fragment from a GB-level directed acyclic graph. Compared to Bitcoin's PoW, Ethash is more focused on the fast IO capabilities of computers.

Different from PoW, PoS does not requires a lot of computing resources to reach consensus on new blocks. PoS selects the creator of the block in a defined manner, which usually depends on the stake of the miners. The more stake, the greater the chance that the miner will be selected as the creator. Current Pos-based blockchain systems use different methods to produce the randomnes in the creator election to ensure system security\cite{gilad2017algorand, kiayias2017ouroboros, buterin2017casper}. However, the current PoS is still very controversial\cite{houy2014will, poelstra2014distributed}, and its security and robustness are not superior to PoW.

\subsection{Smart Contract and Virtual machine}
The concept of smart contract was first proposed by Nick Szabo in 1994\cite{szabo1994smart}, but its ideas have not progressed. Until 2014, as part of the Ethereum smart contract system, smart contract first appeared. In Ethereum, the execution of smart contract is driven by transactions. A smart contract is essentially bytecode. Users write the business logic in a high-level language and then submit and deploy the compiled bytecode to the blockchain. The bytecode that is deployed on the blockchain is called a smart contract. Every smart contract has its own storage space and its own state, and the transfer of state is only determined by the logic of the contract itself.

The bytecode of the smart contract is executed on the blockchain virtual machine. At present, the blockchain virtual machine is divided into Turing-complete virtual machine and Non-Turing-complete virtual machine. The difference between the two is that Turing-complete virtual machine supports the loop statement. Because \textit{halting problem}\cite{turing1937computable} is unsolvable, the Turing-complete blockchain virtual machine needs to introduce a gas mechanism to solve the dead loop problem. Before the smart contract is executed, the user must provide a certain amount of gas. Gas is a scarce resource that is acquired by cryptocurrency purchase or mortgage. The blockchain virtual machine consumes a certain amount of gas to execute bytecode, while different bytecode consume a different amount of gas. The gas consumed by the bytecode shows its computational cost. Under the gas mechanism, even if the logic of the smart contract is an infinite loop, its execution will eventually stop because of the exhaustion of gas. At present, Turing-complete virtual machine is adopted by smart contract platforms such as Ethereum, because only Turing-completeness can support complex decentralized application logic. Non-Turing-complete blockchain virtual machine does not support looping statements, thus avoiding \textit{halting problem}. This type of virtual machine is used by the decentralized transfer platforms, such as Bitcoin, which does not need to support complex logic.

\section{Security Issues in BBGS} \label{BBGSSecurity}
\begin{figure}[htb]
  \centering
  \includegraphics[width=\linewidth]{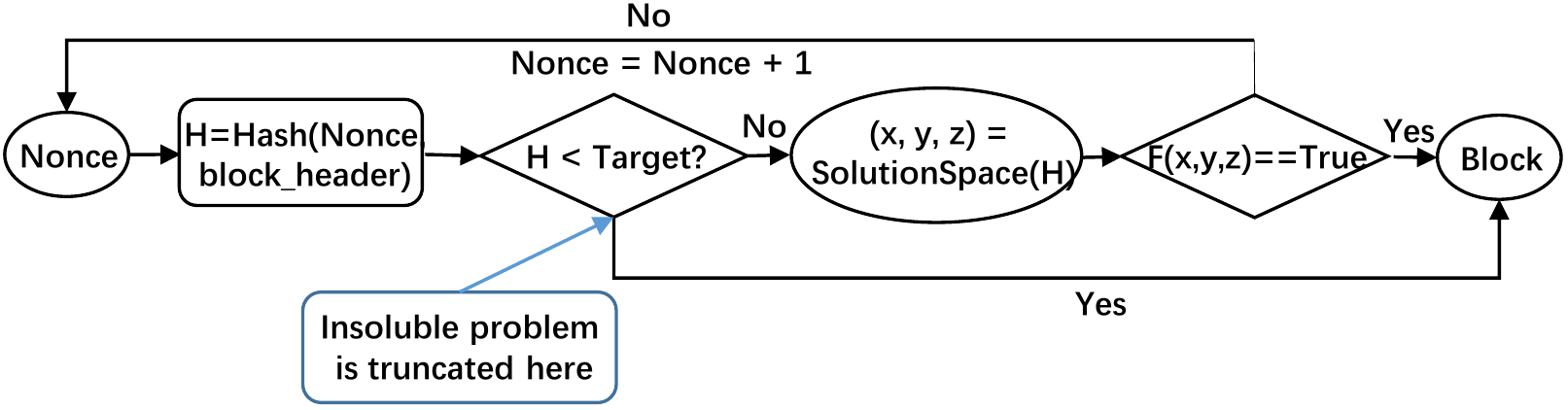}
  \caption{BBGS consensus process}
  \label{twotupleBBGS}
\end{figure}

As shown in Figure \ref{twotupleBBGS}, under \textit{BBGS}, miners constantly perform ${H=Hash(Nonce,block\_header)}$ get a workload ${H}$. If ${H}$ is less than the target, miners can generate a new block; otherwise, they will Map ${H}$ to a potential solution (x, y, z) and then substitute (x, y, z) into function F (the logic of the problem is written inside F) to verify if it is a correct solution to the current problem. If the function F returns True, miners can also generate a new block. \textit{BBGS} can ensure that the system can still generate new blocks by degenerating to the traditional PoW when the problem has no solution. However, this strategy also causes the block-generation time to be affected by the difficulty of the problems, which impairs the availability of the system and makes the difficulty retargeting of the blockchain meaningless. In addition, \textit{BBGS} can lead to serious security issues. Next, we will analyze these issues in detail based on the consensus process in Figure \ref{twotupleBBGS}.

\textit{\textbf{Issue 1}}: An attacker submits a very easy problem whose F(x, y, z) always returns True. When mining, all miners solve the problem instantaneously and then generate and broadcast a new block. At this time, the network will be congested, resulting in a large number of forks which reduces the system availability. In extreme cases, each miner will only work on his own fork and the system's computing power will be diluted to all forks, so most of the computing power will be wasted rather than being used to solve problems. In addition, a large number of forks means that there are multiple ledgers in the system, and the system is extremely unstable.

\begin{figure}[htb]
  \centering
  \includegraphics[width=\linewidth]{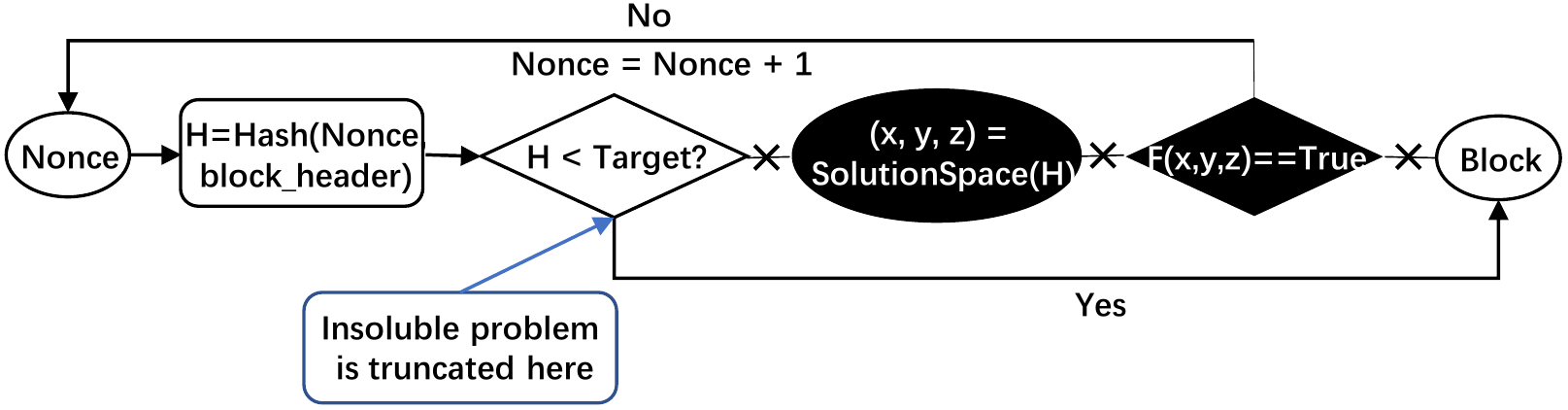}
  \caption{Attack strategy to BBGS}
  \label{twotupleblack}
\end{figure}

\textit{\textbf{Issue 2}}: An attacker submits a problem that takes a long time to calculate but has no solution, that is F(x, y, z) always returns False. The detail of the attack strategy is shown in Figure \ref{twotupleblack}. Here, we analyze the different behaviors of the attacker and honest miners:
\begin{enumerate}
  \item When ${H}{<}{Target}$, there is no difference.
  \item When ${H}{\geq}{Target}$, the behavior of the attacker and honest miners is quite different. The attacker knows that the problem has no solution and calculating the problem is time-consuming, so he will skip the time-consuming operation of the black box in Figure \ref{twotupleblack}. In other words, he does not waste the computing power to calculate the problem. The attacker will directly chooses the new Nonce to start the next round of collision while the honest miners will still calculate the problem before choosing a new Nonce.
\end{enumerate}

Assuming that time of calculating the problem is 1000 times that of one hash calculation, the attacker's equivalent computing power is 1000 times that of a honest miner with the same physical computing power. In this case, an attacker can easily launch a \textit{majority-attack} on blockchain without aggregating a large amount of computing power.

\section{System Design}\label{SystemDesign}
\subsection{Axe blockchain architecture}
\begin{figure*}[htb]
  \centering
  \includegraphics[scale=0.55]{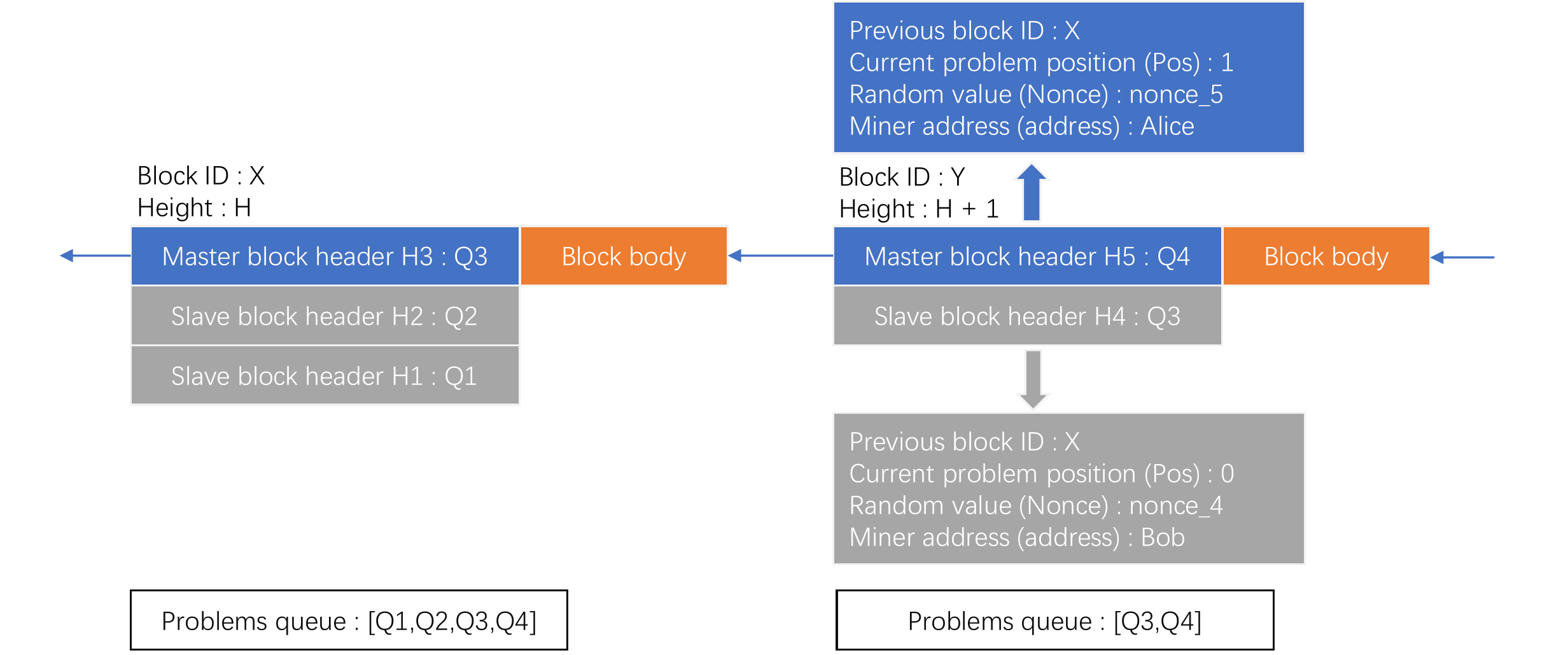}
  \caption{Axe blockchain architecture}
  \label{axeblockchainstruct}
\end{figure*}
Axe blockchain architecture is a new blockchain underlying architecture designed specifically for addressing \textit{Issue 1} of \textit{BBGS}. \textit{Issue 1} is essentially because the block generation is affected by the problem solving, that is, the problem solving becomes one of the decisive factors in the block generation. Since the difficulty of the problem is unknown, the difficulty of mining cannot be adjusted. Figure \ref{axeblockchainstruct} depicts the architecture of the Axe blockchain. Each block in Axe blockchain consists of a block header cluster and a block body. A block header cluster can contain multiple block headers, which are mainly divided into two types: master block header and slave block header. The data structure of Axe blockchain looks like an axe.
\begin{itemize}
  \item \textbf{Master block header (blue flag)}: There must be a unique master block header in the block header cluster. Master block header corresponds to the block body (orange flag) through the Merkel Root. The miner who mines master block header successfully is called the master miner. The master miner is responsible for the packing of the block transactions and the inclusion of the slave block headers. A legal master block header must satisfy the requirement that its workload is less than the target.
  \item \textbf{Slave block header (gray flag)}: The number of slave block header is variable. Slave block header can be mined by slave miners and included into the block header cluster by master miners. A legal slave block header must satisfy the requirement that its corresponding problem is solved. As shown in Figure \ref{axeblockchainstruct}, the problem Q1 corresponding to the slave block header H1 needs to be solved, and the problem Q3 corresponding to the slave block header H4 needs to be solved.
\end{itemize}

Next, we will take the block with H + 1 height as an example to describe the process of block generation in Axe blockchain. First, Alice packages the transactions and constructs a block header. Then, Alice sets the Pos field of the block header to zero and extracts the highest priority problem Q3 from the queue. Alice try to traverse the Nonce to solve Q3. However, when solving Q3, Bob first solves Q3 and broadcasts the corresponding slave block header H4. Alice receives H4 from Bob and verifies that Q3's answer is correct. After successful verification, H4 is included as a slave block header into its own block header cluster. Since Q3 has been solved by Bob, Alice gives up solving Q3 and sets the Pos field to 1 and then extracts the problem Q4 from the queue. When solving Q4, Alice reaches the required workload for the block generation and then inlcudes the block header H5 as a master block header into the block header cluster and broadcasts the entire block. Because Alice took the lead in completing the calculation work of PoW in the blockchain network, Alice's block was verified by other peers and written to the local database. In summary, a legal block header cluster needs to meet the following requirements:

\begin{enumerate}
  \item The Merkel Root on the master block header is equal to that generated by the block body.
  \item The workload of the master block header is less than the difficulty target.
  \item The previous block hash referenced from the slave block headers is the same as that of the master block header, which ensures that the master and slave miners are based on the same blockchain branch for consensus.
  \item The problem corresponding to the slave block header is solved.
  \item The Pos field from slave block header to master block header is starting from zero and continuous. In particular, the Pos field value of the master block header is the largest among all block headers in the block. This condition ensures that the problem is solved in turn and no problem is skipped. This also means that miners can not freely select the problem, ensuring that the computing power of the entire network is calculating the same problem at the same moment.
\end{enumerate}

\begin{figure}[htb]
  \centering
  \includegraphics[width=\linewidth]{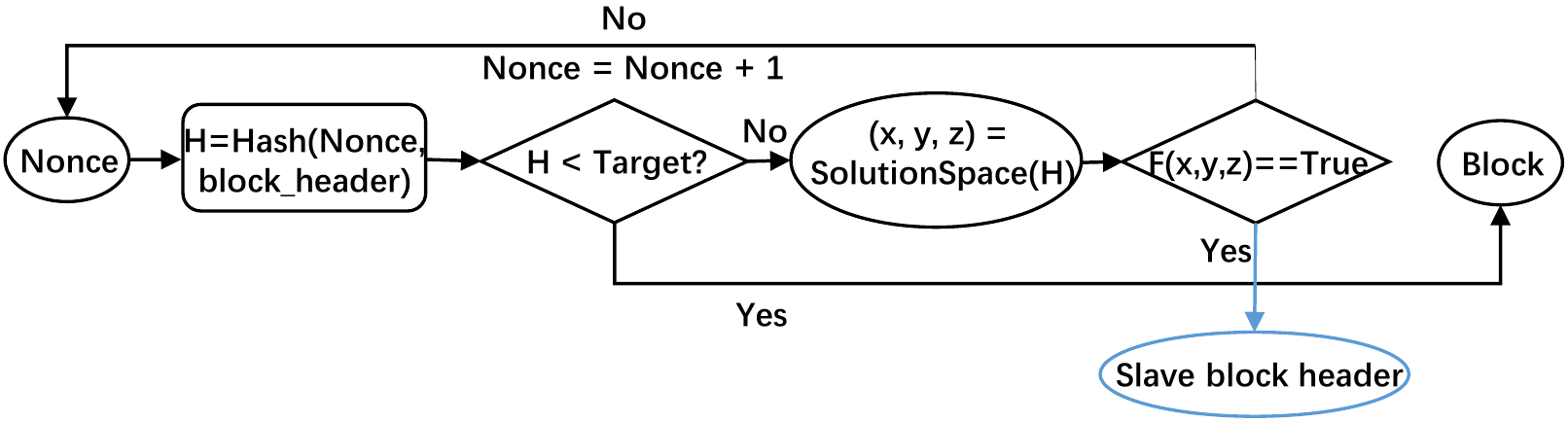}
  \caption{Consensus process in Axe blockchain architecture}
  \label{betterconsensus}
\end{figure}

As shown in Figure \ref{betterconsensus}, in Axe blockchain, a miner only generates a slave block header (the size is small) but not a new block when successfully solving a problem. The new block generation is independent of problem solving and is only controlled by the current difficulty target, so there won't be a lot of forks as described in \textit{Issue 1}. Such a design has great significance for ensuring the stability of block generation rate and the effectiveness of the difficulty retargeting. The size of the slave block header is small so it spreads quickly in the network, thus reducing the possibility of network congestion. At the same time, the number of slave block headers in a block is limited. When encountering potential malicious problems (that is, the network is overwhelmed with many slave block headers), miners can choose not to broadcast and forward the corresponding slave block headers to prevent possible network congestion without affecting the consensus.

\subsection{Problem Definition}
\subsubsection*{\textbf{Problem representation}}
At present, there are two main methods to represent the problem: (1) hard-code, such as the search of prime sequence in Primecoin\cite{king2013primecoin}; (2) representing the problem as a specific format in \cite{oliver2017proposal, chatterjee2019hybrid}, such as reducing NP problems to a type of NP-complete problems. However, these two methods have great drawbacks. The first method can only solve a single problem. This method lacks versatility, practicality and cannot meet the various needs of users. The second method requires users to convert the problem into a standard format and submit the corresponding parameters, which is very unfriendly to users. In addition, standard format conversions takes a lot of work and may introduce higher computational complexity. The second method is less expressive, which makes it difficult to represent problems in complex scenarios.

In this paper, we use Turing-complete high-level language to represent a problem as a smart contract. Therefore, any complex problem can be well represented and the programming of the problem contract is in line with user habits. The design essentially represents all problems as compiled bytecode. The bytecode is the basic computing unit for the blockchain virtual machine. Any problem can be reduced to a combination of basic units, so the amount of work involved in solving any problem can be measured based on the basic unit. In addition, the gas mechanism can limit the running time of the problem contract on the virtual machine, preventing miners from taking too long to verify the correctness of a potential solution to the problem.

\subsubsection*{\textbf{Problem Contract Template}}
A problem contract can be considered as a special type of smart contract. Using smart contract programming language to express common problems has the following advantages: (1) in line with user habits; (2) strong expressive ability; (3) blockchain can be used as a data source; (4) compatible with current smart contract platform tools and workflows. Users can complete the programming of a problem contract by inheriting the provided template and implementing the relevant interface. A problem contract mainly includes tip, deposit, solution space and arithmetic logic. Four public interfaces of the problem contract template can be list as follows:
\begin{itemize}
  \item \textit{getTips}: This interface returns the tip. Users should pay for the system when deploying a problem contract. The total cost inlcudes the tip and deposit. The tip determines the priority of the problem and the deposit is used to pay for the calculation of the problem.
  \item \textit{getSpaceInfo}: This interface returns the type and scope of the solution space. The solution space is an array of n rows and 2 columns, indicating the range of each dimension.
  \item \textit{getTimes}: This interface returns the number of problem solutions that the user requested to find.
  \item \textit{solve}: This interface returns a boolean value, which represents whether the potential solution is correct for the problem. Users can implement the logic code of the problem inside \textit{solve}, describing what is the solution to the problem. \textit{solve} is executed continuously in the virtual machine to solve the problem.
\end{itemize}

\subsubsection*{\textbf{Problem Contract Sample}}\label{ProblemContractSample}
The problem submitted to AxeChain must be Easily-Verifiable, which means that verifying the correctness of a potential solution cannot take too long. Therefore, most NP problems can be submitted to AxeChain because the answer to these problems can be verified in polynomial time. In general, NP problems lack efficient algorithm and finding their solutions takes decades of time on a normal computer, so they are suitable for being submitted to AxeChain, where all computers on a distributed environment attempt to solve a single problem at the same time. In addition, non-linear integer programming problems can also be submitted to AxeChain, most of which were proved to be NP-complete\cite{murty1987some, pardalos1991quadratic}. Even more interesting and exciting is that AxeChain can help humans solve problems in Number Theory, such as solving Diophantine Equation\cite{mordell1969diophantine}. In September 2019, mathematicians Andrew Sutherland and Andrew Booker found a ${(x_1, x_2, x_3)}$ that satisfied ${x_1^3+x_2^3+x_3^3=42}$, causing a sensation in the mathematical world\cite{sum42}.

In addition, AxeChain is well suited for some query work in the blockchain domain. At present, most of the blockchains use key-value databases. Since key-value database does not support the value query very well, the user can only find the potential legal value in the blockchain by traversing the whole database. For example, the blockchain project party selects the active account for \textit{Token Airdrop}\cite{Airdrop}. However, since traversing the whole database is slow and the data on blockchain is updated in real time, the result of the query may be delayed. Similar jobs that filtering data on blockchain can be written as a problem contract submitted to AxeChain. All miners can filter data in parallel, so they can finish the work quickly.

\begin{figure}[htb]
  \centering
  \includegraphics[width=\linewidth]{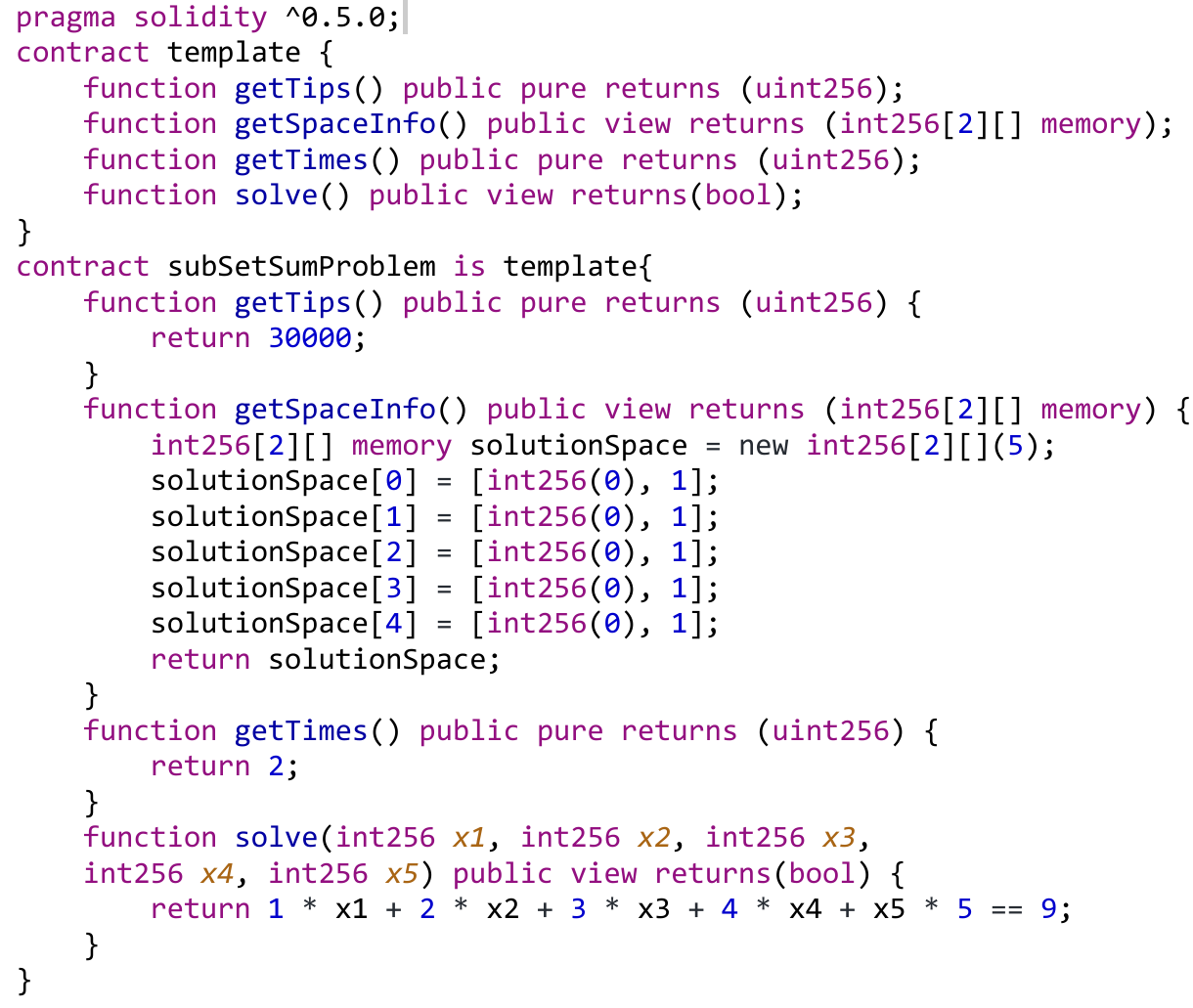}
  \caption{A subset-sum problem contract}
  \label{problemcontract}
\end{figure}

In essence, any problem can be submitted to AxeChain as long as the verification time is not too long (Easily-Verifiable). In addition, AxeChain represents a problem as a smart contract in a high-level language, making the submission of problems more flexible and versatile. Here, we provide a problem contract sample of a classic subset sum problem (a NP problem). As shown in Figure \ref{problemcontract}, this problem contract indicates: Given a set $S = \{1, 2, 3, 4, 5\}$, find a subset of ${S}$ whose sum is 9. It is worth noting that the function getTimes returns 2, which means that the user want the system to find two solutions. More examples of problem contracts can be found in the appendix of this paper.

\subsection{Problem Life Cycle}
In this subsection, we will describe the life cycle of a problem from the process of deployment, consensus, and exit.
\subsubsection*{\textbf{Problem Deployment}}
\begin{figure}[htb]
  \centering
  \includegraphics[width=\linewidth]{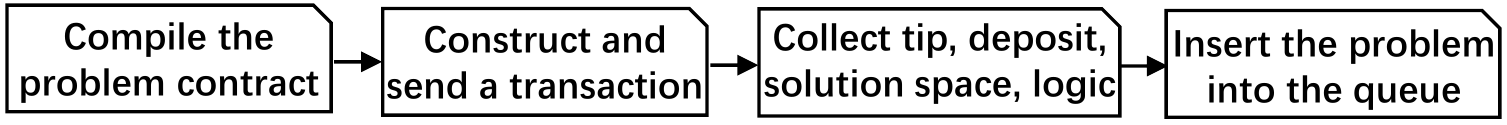}
  \caption{The process of problem contract deployment}
  \label{problemdeploy}
\end{figure}
The process of problem contract deployment is shown in Figure \ref{problemdeploy}. Users compile the problem contract to get the corresponding bytecode, and then use the bytecode to construct a transaction and send the transaction to \textit{System Contract}. \textit{System Contract} is a smart contract that manages problem contracts, that is, maintains a problem queue. The design that deploying a problem by sending a transaction is compatible with current wallet applications.

The cost of deploying a problem is the sum of the deposit and tip. The tip is used to increase the priority of the problem and the deposit is used to reward miners for solving problems. AxeChain's charging model is based on the amount of work involved in calculating the problem, rather than deducting all deposit directly. Such a design mainly takes into account that it is difficult to accurately access the difficulty of the problem and give a fair remuneration. Under AxeChain's charging model, users will be willing to submit problems without worrying about paying too much and miners will not suffer losses due to low remuneration.

After receiving the problem, AxeChain extracts the arithmetic logic (the logic inside the interface \textit{solve}), solution space, deposit and tip of the problem to construct a problem object, and then inserts it into the problem queue.
The problem queue is a priority queue. The priority of the problem is determined by the tip and insertion time, and is not affected by the deposit. It is worth noting that the tip will be destroyed directly without being given to miners. This design is to prevent miners from receiving problems with high tip only. The problem priority ${socre}$ can be calculated as:
\begin{equation}
  score(i, tip) = tip\times{k}^{-i}
\end{equation}

The tip is given by the user, and i is the order in which the problem is inserted in the current queue. ${i = 0}$ means that the problem is the earliest insertion of the current queue. ${k}$ is a constant greater than 1, used to weigh the impact of ${tip}$ and ${i}$ on priority. Higher tip and earlier insertion mean higher priority. The relationship between the ${tip}$ and ${i}$ is shown in Equation \ref{equation2}. Equation \ref{equation2} shows that if the tip of a problem is k times that of the previous one, the problem has the same priority as the previous one. There is a default problem in the queue (such as looking for prime sequence) and its priority is the lowest in the queue. The default problem is to prevent the queue from being empty.
\begin{equation}
  score(i, tip) {\equiv} score(i + 1, k{\times}tip) \label{equation2}
\end{equation}
\subsubsection*{\textbf{Problem Consensus}}
\begin{figure}[htb]
  \centering
  \includegraphics[width=\linewidth]{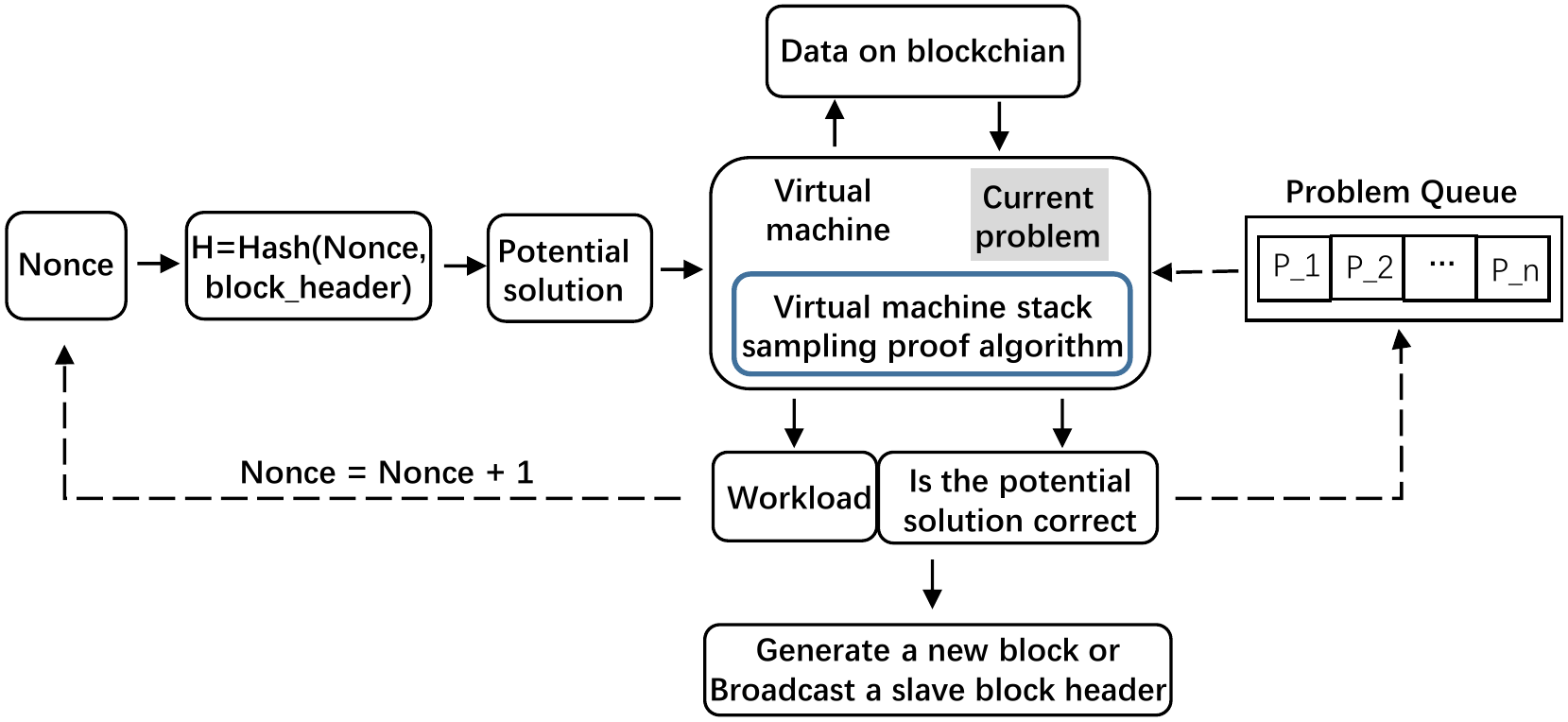}
  \caption{The process of problem consensus}
  \label{problemconsensus}
\end{figure}

As shown in Figure \ref{problemconsensus}, when mining, miners first retrieve the highest priority problem from the problem queue and load the bytecode of the problem into the Turing-complete stateful blockchain virtual machine. That is to build a virtual machine based on the latest state and data of blockchain as a container to run bytecode to solve the problem. The virtual machine can only read data from system. Then, miners traverse the Nonce field and constantly perform ${H=Hash(Nonce,block\_header)}$ to get a hash ${H}$. Furthermore, miners map ${H}$ into the problem solution space to get a potential solution. Given the problem solution space ${[a_1, b_1]\times[a_2, b_2]\times\ldots\times[a_n, b_n]}$ and ${H}$, the potential solution ${s}$ can be calculated by the following equations:
\begin{equation}
  \begin{aligned}
  s_i &= a_i + H_i\,mod\,(b_i - a_i + 1) \\
  H_{i + 1} &= H_i / (b_i - a_i + 1) \\
  s &= (s_1, s_2, \ldots, s_n)
  \end{aligned}
\end{equation}

The blockchain virtual machine runs the bytecode (the interface \textit{solve}) of the problem with s as input, and it can acquire data from the blockchain. During the virtual machine running process, miners not only try to solve the problem, but also generate a final workload through the \textbf{Virtual machine stack sampling proof algorithm}. If a miner find a correct solution, it can broadcast a slave block header and if it find a workload smaller than the target, it can generate a new block. The detail of one-round consensus is shown in Algorithm \ref{problemconsensusprotocol}, where ${min\_work}$ is used to measure the amount of work involved in calculating a problem (the smaller ${min\_work}$ is, the problem is more difficult and the more work).
\begin{breakablealgorithm}
  \caption{Problem consensus protocol}
  \label{problemconsensusprotocol}
  \begin{algorithmic}[1]
    \State construct ${header_0}$ 
    \State ${current\_problem}{\gets}{problem\_queue[0]}$
    \State ${pos}{\gets}{0}$ \&\& ${slave\_headers}{\gets}{\phi}$
    \State ${nonce}{\gets}{union(0, max(NONCE) - 1)}$  \Comment{initial randomly}
    \State ${solved}{\gets}{False}$ \Comment{it represents whether the current problem is solved}
    \State ${min\_work}{\gets}{max(WORK)}$ \Comment{it records the minimum workload found in solving ${current\_problem}$}
    \State ${min\_nonce}{\gets}{nonce}$ \Comment{it corresponds to ${min\_work}$}
    \State ${work}{\gets}{max(WORK)}$  \Comment{initial}
    \While {${work > target}$}
      \If{${solved}$} // if the current problem is solved
        \State // construct a slave block header
        \State ${header_{slave}}{\gets}{pack(header_0, pos, nonce, min\_nonce)}$
        \State // include ${header_{slave}}$ into ${slave\_headers}$
        \State ${pos}{\gets}{pos + 1}$ \&\& ${slave\_headers}{\gets}{header_{slave}}$
        \State broadcast ${slave\_headers}$
        \State ${current\_problem}{\gets}{problem\_queue[pos]}$ 
        \State ${min\_work}{\gets}{max(WORK)}$ 
        \State ${min\_nonce}{\gets}{nonce}$
      \EndIf
      \If{${header_{slave}}{\gets}{network}$ \&\& ${valid(header_{slave})}$}
      \State ${pos}{\gets}{pos + 1}$ \&\& ${slave\_headers}{\gets}{header_{slave}}$
      \State ${current\_problem}{\gets}{problem\_queue[pos]}$
      \State ${min\_work}{\gets}{max(WORK)}$
      \State broadcast ${header_{slave}}$
      \EndIf
      \If{${block_{new}}{\gets}{network}$ \&\& ${valid(block_{new})}$}
      \State ${blockchain}{\gets}{block_{new}}$ \&\& broadcast ${block_{new}}$
      \State \Return
      \EndIf
      \State ${nonce}{\gets}{nonce + 1}$
      \State ${work, solved}{\gets}$\textbf{stack\_proof}${(header_0, nonce, }$${current\_problem)}$
      \If{${work < min\_work}$}
      \State ${min\_work}{\gets}{work}$ \&\& ${min\_nonce}{\gets}{nonce}$
      \EndIf
    \EndWhile
    \State // construct a master block header
    \State ${master\_header}{\gets}{pack(header_0, pos, nonce)}$
    \State ${block}{\gets}{pack(master\_header, slave\_headers, block\_body)}$
    \State ${blockchain}{\gets}{block}$ \&\& broadcast ${block}$
    \State \Return
  \end{algorithmic}
\end{breakablealgorithm}

\subsubsection*{\textbf{Problem Exit}}
When the current problem is successfully solved and AxeChain has paid miners for solving the problem, the problem is removed and all remaining deposit are returned to the user. When the problem is not resolved and there is still enough deposit to support the next round of solving after paying for miners, the problem is reinserted into the queue; otherwise, the problem is removed and the remaining deposit is returned.

\section{Stack sampling proof algorithm}\label{samplingproof}
\subsection{Virtual machine stack sampling proof}
Virtual machine stack sampling proof algorithm is designed specifically for addressing \textit{Issue 2} of \textit{BBGS}. \textit{Issue 2} is essentially caused by the fact that the workload generation is not coupled to the problem solving, that is, miners can skip the calculation of the problem and generate a workload. In this paper, we choose stack-based virtual machine as a container for running problem contracts. When bytecode is executed in the virtual machine, the impact of the corresponding operation is mainly reflected on the top of the stack. We sample the top elements of the stack and aggregates the sampling results to get a digest ${stack\_digest}$ of the contract running state. The input to the problem contract is determined by the hash ${H}$ (${H=Hash(Nonce,block\_header)}$), so ${stack\_digest}$ is uniform and random. Different Nonce corresponds to different ${H}$, different ${H}$ corresponds to different input to the problem contract, so the stack contents and ${stack\_digest}$ is different. Finally, ${H}$ is concatenated with ${stack\_digest}$ to generate the final workload. In this way, the final workload generation and problem solving can be tightly coupled. In order to get ${stack\_digest}$, the miner should run the problem bytecode according to the input to get the stack contents.
\subsubsection*{\textbf{Overview}} The stack sampling proof algorithm has the following basic requirements, which essentially make solving problem as a superior strategy for miners.
\begin{enumerate}
  \item Miners cannot predict the sampling location. The stack should be sampled at different location for different ${H}$ (different Nonce).
  \item Stack sampling can only performed serially with the running of bytecode and cannot be done in parallel, and the number of samplings is sufficient to get a more comprehensive ${stack\_digest}$.
  \item 
  Single sampling should takes less time than running a problem contract. If a single sampling takes much more time than running problem contract, sampling is meaningless.
\end{enumerate}

Inspired by Ethereum Ethash\cite{ethash}, Virtual machine stack sampling proof algorithm uses cryptographic hash algorithm SHA-3 at both ends to ensure security, and uses FNV to aggregate the sampling results of the stack to get  ${stack\_digest}$. FNV is a classic lightweight non-cryptographic hash algorithm. In each round of collision, miners can only perform ${H=SHA3(Nonce,block\_header)}$ and map ${H}$ to a potential solution to get an input to the problem contract. This design ensures that miners cannot predict the input to the problem contract, so miners cannot predict ${stack\_digest}$. Since the final workload is the result of ${SHA3(H||stack\_digest)}$, miners cannot get a final workload without calculating the problem. As long as ${stack\_digest}$ generated by the stack sampling and aggregating is sufficiently discrete making it difficult for miners to predict ${stack\_digest}$, miners have little chance of evil and the system security can be ensured. 

\subsubsection*{\textbf{Generation of final workload}} This paper uses 32 bytes of SHA-3 and 8 bytes of FNV. The bit width of current blockchain virtual machine is 256 (32 bytes). As shown in Equation \ref{initialsnapshot}, the initial snapshot of stack sampling is obtained by aggregating ${H}$ through FNV, which costs three FNVs. The hash ${H}$ is grouped by 8 bytes during the aggregation process, for example, ${H[0]}$ represents the first 8 bytes of the hash.

\begin{equation}
  snapshot_0 = FNV(FNV(H[0],H[1]),FNV(H[2],H[3])) \label{initialsnapshot}
\end{equation}

The stack changes when running problem contract. The elements at the top of the stack are most closely related to the current operation, so stack sampling should focus on the top elements of the stack. If there are not enough elements in the stack, we can supplement it with H. This paper uses FNV to aggregate the four elements at the top of the stack to update the snapshot. The new snapshot is obtained by sequentially performing FNV in bytes, and finally merging the intermediate results. The aggregation is started from the initial snapshot to prevent the attacker from designing the problem contract, and calculating a part of the intermediate results in advance. Since ${H}$ is calculated by SHA-3, the attacker cannot predict the initial snapshot and calculate the intermediate results in advance. More details are shown in Algorithm \ref{snapshotupdate}.

\begin{algorithm}
  \caption{Snapshot update}
  \label{snapshotupdate}
  \begin{algorithmic}[1]
    \Require
      Stack set: \{${stack_0, stack_1, stack_2, stack_3}$\}, current snopshot: ${snapshot_{i-1}}$
    \Ensure
      A new snapshot: ${{snapshot}_{i}}$
    \State ${T}{\gets}{\emptyset}$
    \For{$stack$ \textbf{in} \{${stack_0, stack_1, stack_2, stack_3}$\}}
    \For{$j = 1 \to 32$}
    \State ${snapshot}{\gets}{FNV(snapshot, uint64(stack[j]))}$
    \EndFor
    \State ${T.append(snapshot)}$
    \EndFor
    \State ${{snapshot}_{i} = FNV(FNV(T[0],T[1]),FNV(T[2],T[3]))}$
    \State ${{snapshot}_{i} = FNV({snapshot}_{i}, {snapshot}_{i-1})}$
    \State \Return${{snapshot}_{i}}$
  \end{algorithmic}
\end{algorithm}

In order to ensure that the sampling location are not predictable, this paper uses the previous snapshot to update the sampling location for current sampling. Given that the snapshot after ${i-1}$ samplings is ${snapshot_{i-1}}$ and the stack is sampled every time the problem contract consumes ${v}$ gas, the location for next sampling can be calculated as
\begin{equation}
  next\_loction = (i-1)v + snapshot_{i-1}\,mod\,v
\end{equation}

The sampling method of this paper ensures that the sampling location distribution and the stack contents are different for different potential solutions. At the same time, if ${snapshot_{i-1}}$ is not calculated, the next sampling location is unpredictable, thus ensuring that the stack sampling runs serially.

Finally, the workload ${work_0}$ can be calculated by Equation \ref{sha3work}. ${state\_digest}$ is the last snapshot of the stack sampling. Since ${state\_digest}$ is obtained by the non-cryptographic hash algorithm FNV, its distribution will be slightly uneven. Therefore, we use SHA-3 to guarantee the uniformity distribution of ${work_0}$. However, ${work_0}$ cannot be directly compared to difficulty target. The cost of generating a workload is different for different problem contracts, so a uniform equivalent workload is required. Assume that the gas consumed by one SHA-3 is ${k}$ times one FNV. For a problem contract with ${n_{sample}}$ sampling locations, it takes ${n_{fnv}}$ FNVs to generate a workload, including ${2k}$ FNVs for SHA-3 at both ends, 3 FNVs for the initial snapshot and ${(32{\times}4+3+1){\times}{n_{sample}}}$ FNVs for ${n_{sample}}$ samplings (in Algorithm \ref{snapshotupdate}). The final (equivalent) workload $work_{final}$, which is used to compare against the target, can be calculated by Equation \ref{finalWork}. If $work_{final} < target$, the miner can generate a new block and get reward.

\begin{equation}
  work_0 = SHA3(H||state\_digest) \label{sha3work}
\end{equation}

\begin{equation}
  \begin{aligned}
    n_{fnv}&= 2k + 3 + {(32{\times}4+3+1){\times}{n_{sample}}}\\
    work_{final} &= work_{equivalent} = work_0 / n_{fnv}
  \end{aligned}
  \label{finalWork}
\end{equation}

\subsection{System security analysis}
Here we assume an extreme case: the attacker carefully designs a problem contract whose stack contents do not change for different inputs at runtime. Therefore, the attacker can obtain the stack contents in advance without running the problem bytecode, while the honesty still needs to run the problem bytecode. In this case, the attacker gains a relative advantage. However, the attacker still needs to calculate ${state\_digest}$ through stack sampling algorithm. Even if the stack contents are the same, ${state\_digest}$ will vary depending on ${H}$ (the input to the initial snapshot). Therefore, miners still need to do stack sampling but does not need to run the bytecode.

\subsubsection*{\textbf{Utilization radio}}
The computing power of AxeChain is used not only to execute bytecode for solving problems, but also to stack sampling and SHA-3. The utilization ratio of computing power ${\eta}$ is used to measure the proportion of the power that is actually used for problem solving, and its definition is shown in Equation \ref{Utilizationradio}. ${{g}_{total}}$ is the gas consumed to run the problem contract and gas is used to measure the cost of the calculation. ${{n}_{fnv}}$ represents the number of FNVs in the stack sampling and SHA-3, and ${g_{fnv}}$ represents the gas consumed by one FNV. The larger ${{n}_{fnv}}$${g_{fnv}}$, the more computing resources the stack sampling consumes. The smaller ${{n}_{fnv}}$${g_{fnv}}$, the higher the utilization radio of computing power, that is, only a small amount of computing power is used for stack sampling and SHA-3, and most of the computing power is used for problem calculation. From a macro perspective, the computational power that is not used to solve the problem is used to maintain the security of blockchain.
\begin{equation}
  {\eta}=\frac{g_{total}}{n_{fnv}g_{fnv}+g_{total}} \label{Utilizationradio}
\end{equation}
\subsubsection*{\textbf{Security model}}
Assume that the proportion of honest miners' and attackers' computing power to the whole network is ${p}$ and ${q}$, where ${p + q = 1}$. Different from the traditional PoW, the mining difficulty of honest miners and attackers may be different in AxeChain. An attacker can submit a well-constructed problem contract, which allow him to mine without running the problem bytecode. Compared with honest drivers, attackers do not need to spend ${g_{total}}$ gas to run bytecode. The larger ${g_{total}}$ is, the larger ${n_{sample}}$ and ${n_{fnv}}$ are. We can adjust the sampling frequency to adjust the gap between ${g_{total}}$ and ${n_{fnv}g_{fnv}}$, that is ${\eta}$. In the case of the same physical computing power, the attacker can generate a block faster than the honest miners. The ratio of block generation rate between the attackers and honest miners can be calculated as:
\begin{equation}
  {\delta} = \frac{q/n_{fnv}g_{fnv}}{p/(n_{fnv}g_{fnv} + g_{total})} = \frac{q}{p}\frac{1}{1-{\eta}} \label{delta}
\end{equation}

${q/n_{fnv}g_{fnv}}$ represents the attackers' equivalent computing power while ${p/(n_{fnv}g_{fnv} + g_{total})}$ represents the miners' equivalent computing power. In addition, ${\eta}$ represents the utilization ratio of computing power. According to \textit{gambler's ruin problem}\cite{harik1999gambler}, in the case of being behind ${z}$ blocks, the probability that attackers can successfully catch up with honest miners is:
\begin{equation}
Q_z({\delta})=\left\{
    \begin{aligned}
    &{\delta}^{z}  & \ {\delta} < 1 \ and z > 0\\
    &1 &\ {\delta} \geq \ or \ z \leq 0
    \end{aligned}
    \right.
\end{equation}
Assume that the blockchain system is \textit{${y}$ confirmations}. \textit{${y}$ confirmations} means that a transaction is considered almost irreversible when it is supported by ${y}$ blocks. In Bitcoin, ${y}$ is equal to 6. When \textit{${y}$ confirmations} is observed on the main chain, the number of blocks generated by the attacker in the darkness satisfies the \textit{Poisson distribution} with ${{\lambda}={\delta}y}$ and we record it as ${P(x)}$. The probability of a successful attack is:
\begin{equation}
  \begin{aligned}
    P(an\ attack,\ y\ confirmations) &= \sum_{x=0}^{+{\infty}}P(x)Q_z({y-x})\\
      &= 1 - \sum_{x=0}^{y}\frac{{\lambda}^x{e}^{-\lambda}}{x!}(1-\delta^{y-x})
  \end{aligned}
  \label{attackprobability}
\end{equation}
It can be seen from Equation \ref{attackprobability} that when ${{\delta}<1}$, the probability of a successful attack decreases exponentially with the increase of ${y}$. In other words, the system is robust as long as ${{\delta}<1}$. According to ${{\delta}<1}$, ${p+q=1}$ and Equation \ref{delta}, we can deduce that the radio of attackers' computing power ${q}$ has the following constraint:
\begin{equation}
  q<1-\frac{1}{2-{\eta}}
\end{equation}

\begin{figure}[htb]
  \centering
  \includegraphics[scale=0.6]{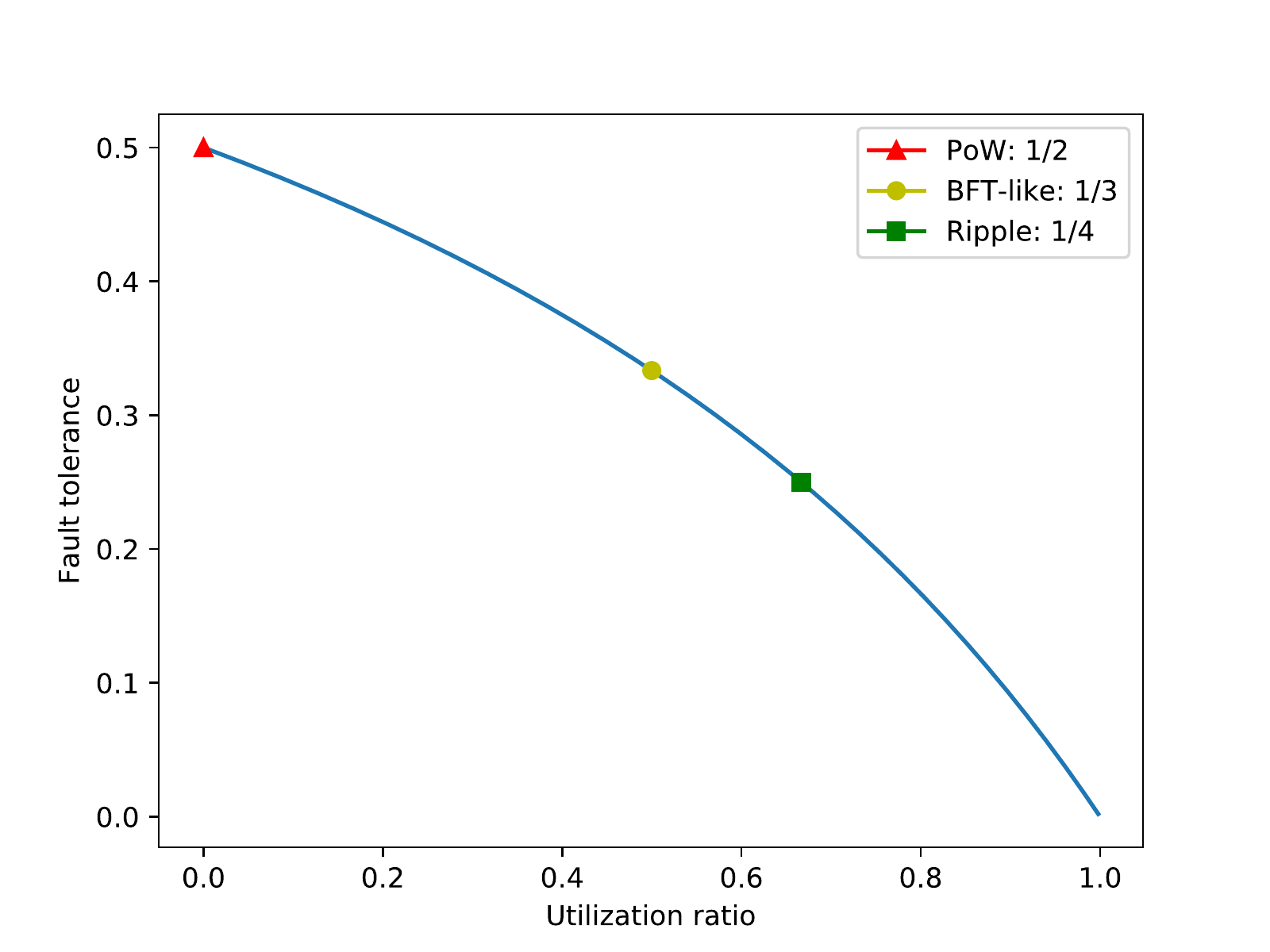}
  \caption{The relationship between utilization radio and system fault tolerance}
  \label{radiocurve}
\end{figure}

By plotting the function ${f(x)=1-\frac{1}{2-x}}$, we can get an balance curve of power utilization radio and system security (fault tolerance), which is shown in Figure \ref{radiocurve}. Overall, system security slowly decreases as the utilization radio of computing power increases. It is difficult to reach a consensus on a system which aimed to solve general problems submitted by users. Therefore, the security of the system is unlikely to be ensured when the power utilization radio reaches 100\%. The system needs a part of the computing power to maintain the security of the blockchain system, and \textbf{Virtual machine stack sampling proof algorithm} is the means used in this paper to maintain system security.

As shown in Figure \ref{radiocurve}, as long as slightly sacrificing the system security factor from 1/2 of PoW to that of the BFT-like algorithm with 1/3 of fault tolerance\cite{castro1999practical, kotla2007zyzzyva, yin2018hotstuff}, the system can achieve 50\% of power utilization; slightly sacrificing system security factor to that of the ripple algorithm with 1/4 of fault tolerance\cite{chase2018analysis}, the system can achieve 66.7\% of power utilization. At the peak, Bitcoin consumed 60 billion kWh/year\cite{bitcoinenergyconsumption}. Taking the computing power of Bitcoin as the benchmark and 1/3 as the security factor, AxeChain will consum 30 billion kWh/year to solve the user's practical problems. It is undeniable that the blockchain system at this time is not only a secure decentralized ledger, but also a decentralized computing platform with strong computing power.

\section{Incentive model} \label{incentivemodel}
A reasonable incentive model is very important to the blockchain, which needs to balance the interests of all parties involved. In the existing researches\cite{oliver2017proposal, chatterjee2019hybrid, amar2019incentive}, the reward for solving problems is completely subjectively determined by users. In most cases, it is difficult for a user to give a reasonable reward because they do not know the difficulty of the problem. Users worry about paying too much, while miners worry about too little income. The incentive model in this paper will assign reward to miners according to the amount of work exactly involved in calculating the problem and the extra rewards will be returned to users. Consequently, it can avoid the loss of both the users and miners.

The incentive mode refers to the deduction of users' deposit and the way miners receives the reward when a new block is generated. Assuming that the probability of reaching the block generation requirement in a single collision is ${p_0}$, then the expected number of collisions to generate a new block is ${1/p_0}$. The benchmark excitation ${W}$ is used to measure the calculation cost of ${1/p_0}$ collisions. ${W}$ can be estimated by ${1/p_0}$ and the average gas consumed for per collision based on different problem contracts. In AxeChain, a block header cluster consists of a master block header and ${k}$ slave block headers. All block headers correspond to the problems ${\{Q_{n-k}, Q_{n-k+1}, {\ldots}, Q_{n-1}, Q_{n}\}}$ and users’ deposit ${\{Deposit_{n-k}, Deposit_{n-k+1}, {\ldots}, Deposit_{n-1}, Deposit_{n}\}}$.

First, the system assigns reward for the master block header. The system deducts W from ${Deposit_{n}}$ as a reward for the master miner. The master miner receives the slave block headers corresponding to ${Q_{n-k}, Q_{n-k+1}, {\ldots}, Q_{n-1}}$, indicating that until the master miner receives the slave block corresponding to ${Q_{n-1}}$, the whole network still has not found a valid workload through collision. Hash collision is memoryless, so the expected time for the entire network to find a valid workload is irrelevant to the hash collision before ${Q_{n}}$. When solving ${Q_{n}}$, the expected number of collisions to find a valid workload is equal to that required by block generation, corresponding to the benchmark excitation ${W}$. The derivation is shown in Equation \ref{QN}. 
\begin{equation}
  E(\mu_{Q_m}) = \sum_{x=1}^{+\infty}x(1-p_0)^{x-1}p_0 = 1 / p_0 \label{QN}
\end{equation}

Next, the system assigns reward for the slave block headers. The expected number of collisions for all slave block headers also tends to that required by block generation. Its derivation is shown in Equation \ref{QNK}. It can be understood here that if the expected number of collisions required to solve ${Q_{n-k}}$ is small, then there will be more time to solve ${Q_{n-k+1}}$, and so on. Therefore, the corresponding calculation cost for ${Q_{n-k}, Q_{n-k+1}, {\ldots}, Q_{n-1}}$ is W.
\begin{equation}
  \begin{aligned}
    & E(\mu_{{Q_{n-k}, {\ldots}, Q_{n-1}}  } | no\ block\ generation \ before \ Q_n) \\
    = & \sum_{x=1}^{+\infty}x\frac{(1-p_0)^x}{\sum_{y=1}^{+\infty}(1-p_0)^y} = 1 / p_0
  \end{aligned}
  \label{QNK}
\end{equation}

It is determined that corresponding calculation cost for ${Q_{n-k}, Q_{n-k+1}, {\ldots}, Q_{n-1}}$ is W, and the system needs to determine the calculation cost for each problem and deduct it from the corresponding deposit. Equation \ref{COST} shows the calculation cost ${Cost_{Deposit_{n-i}}}$ for a single problem, including variable portion ${0.5W\psi_{n-i}}$ and fixed portion ${0.5W/k}$. k represents the number of block slave headers ${1 \leq i \leq k}$. ${\psi_{n-i}}$ reflects the proportion of each problem in variable portion ${0.5W\psi_{n-i}}$. ${min\_work_{n-i}}$ indicates the minimum workload found in solving ${Q_{n-i}}$ from the slave miner (in Algorithm \ref{problemconsensusprotocol}), reflecting the difficulty of ${Q_{n-i}}$. If ${Q_{n-i}}$ is more difficult, ${min\_work_{n-i}}$ will be smaller. ${min\_work_{n-i}}$ is provided from the slave miner. Although it is not equal to the minimum workload found when the whole network solves ${Q_{n-i}}$, it reflects the difficulty of ${Q_{n-i}}$ to some extent. The fixed portion ${0.5W/k}$ is to ensure that the slave miner would broadcast the corresponding slave block header when it successfully solved a problem. If ${Cost_{Deposit_{n-i}}}$ is calculated based only on ${\psi_{n-i}}$, that is ${Cost_{Deposit_{n-i}}} = W\psi_{n-i}$, the following may occur: A lucky miner solves a problem with only one collision, but it will not broadcast a slave block header because ${min\_work_{n-i}}$ is too large. Therefore, a design that requires a fixed portion ${0.5W\psi_{n-i}}$ is necessary.
\begin{equation}
  \begin{aligned}
    \psi_{n-i} &= \frac{\frac{1}{min\_work_{n-i}}}{\sum_{i=1}^{k}\frac{1}{min\_work_{n-i}}} \\
    Cost_{Deposit_{n-i}} &= \frac{W}{2}(\frac{1}{k}+{\psi}_{n-i})
  \end{aligned}
  \label{COST}
\end{equation}

In addition, when a miner is both the master miner and slave miner, it is unable to obtain the reward corresponding to the slave block header. Otherwise, when a miner receives a slave block header (corresponding to problem Q), the miner thinks that it is also possible for him to solve Q and get a reward and then refuse to include the slave block header. Therefore, If there are some slave block headers that are generated by the master miner, the corresponding reward should be deducted from the corresponding user deposit and then be destroyed.

The incentive model based on the amount of work exactly involved in calculating the problem allows the user not to worry about giving too much deposit, but also to ensure that miners can obtain a reasonable income. At the same time, this incentive model is based on the fact that miners not only provide a secure decentralized system, but also solve meaningful practical problems constantly, so it can provide more value endorsement for the cryptocurrency it creates.

\section{Experimental Results} \label{ExperimentalResults}
We implement all components of AxeChain based on Ethereum Client, Geth\cite{geth}. Our implementation mainly modifies the data layer, conseneus layer, incentive layer, and contract layer of Geth. In order to evaluate AxeChain, we mainly conducted two experiments, \textbf{Security Experiment} and \textbf{Problem Solving}. The former is to verify that stack sampling proof can ensure system security on different problem contracts; the latter is to get the relationship between the problem difficulty, deposit, and probability of successful solving to evaluate the availability and feasibility of AxeChain.
\subsection{Security Experiment}
The problems raised by users are innumerable, but the computational requirements of the problem can be essentially decomposed into the basic operation units including addition, subtraction, multiplication, division, and integer modulo. These arithmetic unit constitutes five basic modes for the change of the stack contents. For comparison, we also studied the mode in which the stack contents are constant. In this subsection, we conduct an experimental analysis of the nature and distribution of ${state\_digest}$ (64bit) obtained from the above six modes to verify that the stack sampling can ensure system security on any problem contract.

Algorithm \ref{testfun} shows the test function for each mode, where the input parameter ${X}$ is derived from ${H}$ (${H=Hash(Nonce,block\_header)}$). After the test function performs a basic operation unit on the input parameters $x_i$ in turn, and compares result with the threshold to determine whether to generate a slave block header. The test function can be represented as a specific problem contract and loaded into the virtual machine for execution. We use 12 unsigned-64bit integers as ${X=\{{x_1, x_2, {\ldots}, x_k}\}}$ (ie ${k=12}$) and let ${n=100}$ (ie 8-9 operations for each ${x_k}$ in one execution of test function). For each mode, we generated ${10^6}$ uniformly distributed ${X}$ through SHA3 and set four different sampling frequencies (10\%, 20\%, 30\%, 40\%), and finally collected all the stack digests, a total of ${24{\times}10^6}$ digests for ${24}$ combinations (${4}$ frequencies ${\times}$ ${6}$ modes). Assuming a contract consumes 1000 gas for one exection, then 10\% sampling frequency means that the stack is sampled every 100 gas.

\begin{algorithm}
  \caption{Test function}
  \label{testfun}
  \begin{algorithmic}[1]
    \Require
      Operation type: ${Op}$, input parameter: ${X=\{{x_1, x_2, {\ldots}, x_k}\}}$, number of operations: ${n}$, threshold of solution: ${r_0}$
    \Ensure
      A boolean value
    \State ${result}{\gets}{\emptyset}$
    \For{$j = 0 \to n-1$}
      \State ${result}{\gets}{Op(result, x_{(i\ mod \ k)+1})}$
    \EndFor
    \State \Return${{result}<{r_0}}$
  \end{algorithmic}
\end{algorithm}

\begin{figure}[htb]
  \centering
  \includegraphics[width=\linewidth]{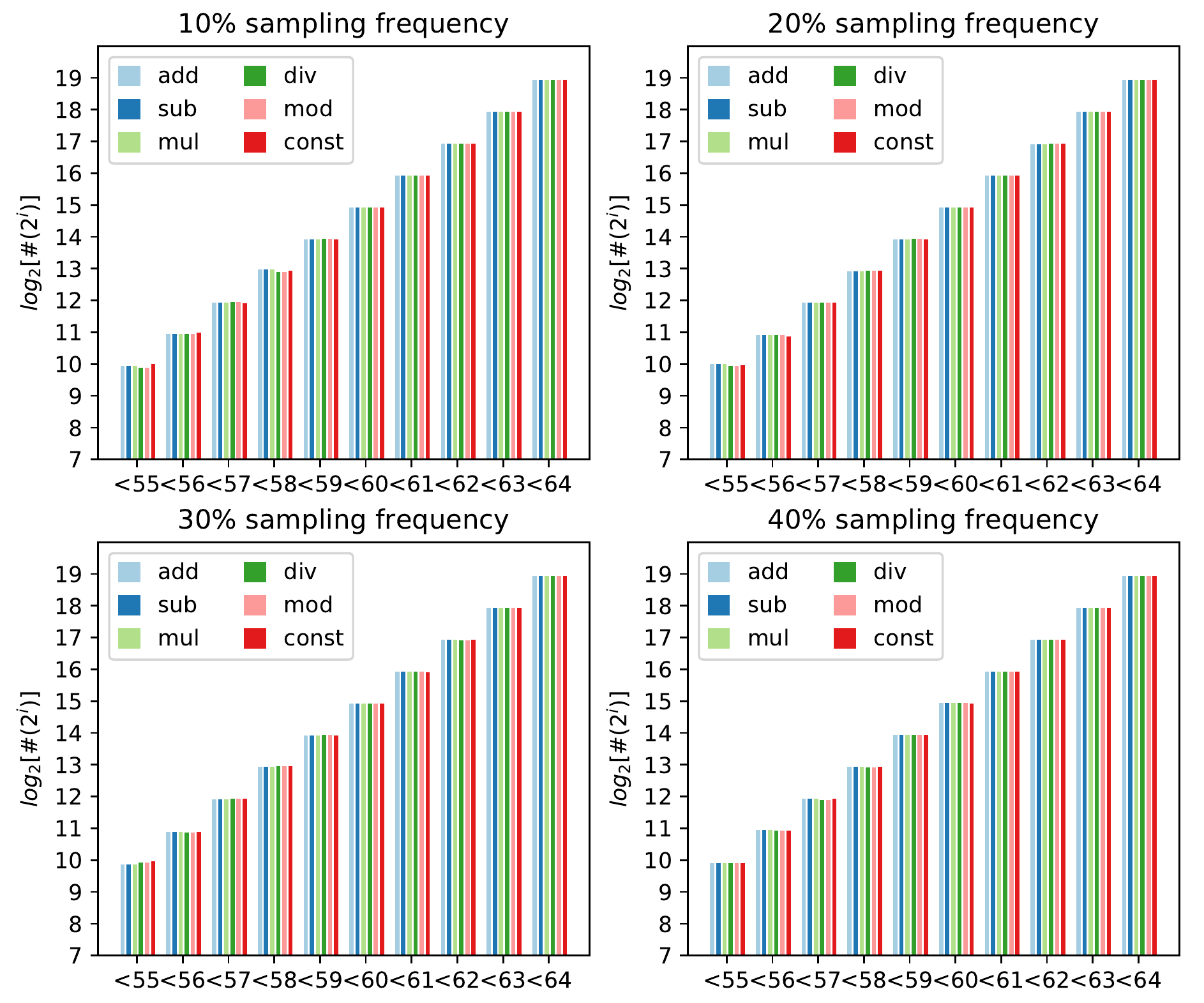}
  \caption{Distribution of ${state\_digest}$ in different modes}
  \label{stackdigest}
\end{figure}

We performed a statistical analysis on all the stack digests, and found that these ${24{\times}10^6}$ digests did not collide (that is, they are different from each other). It is worth noting that even if the stack content is constant, ${state\_digest}$ for different X are still different. Therefore, even if the attacker can calculate the stack contents in advance, it can not skip the calculation of the digest to calculate a final workload. In addition, we also analyze the distribution of ${state\_digest}$ obtained from all modes at different sampling frequencies, and then show the uniformity distribution of ${state\_digest}$ in 64-bit binary space, which is shown in Figure \ref{stackdigest}. 

For every combination, we counted the frequency of ${state\_digest}$ in the range ${[0,2^i]}$ as ${\#(2^i)}$, and then we plotted with ${i}$ as abscissa and ${log_2[\#(2^i)]}$ as ordinate, which is shown in Figure \ref{stackdigest}. Ideally, the distribution law of the frequency satisfies ${\#(2^i)=2{\times}\#(2^{i-1})}$, that is, equivalent to Equation \ref{frequency}. As shown in Figure \ref{stackdigest}, the distribution of ${state\_digest}$ under 24 combinations satisfies Equation \ref{frequency}, showing a linear growth trend with a growth rate close to 1. It shows that ${state\_digest}$ is sufficiently discrete and uniform in different combinations, so that it is difficult for an attacker to design a problem contract to attack the system by exploiting the deviation of ${state\_digest}$ uniformity. 

\begin{equation}
  log_2[\#(2^i)]=log_2[\#(2^{i-1})] + 1 \label{frequency}
\end{equation}

\subsection{Problem Solving}
In this Experiment, we deploy 32 nodes locally and submitted a series of problems to calculate the frequency that each set of problems of different difficulty will be successfully solved under different reward. We simulate networks of 32 local nodes by 32 machines, each of which runs a AxeChain instance with four threads for mining (occupying 50\% CPU utilization). Each machine has an Inter(R) Core(TM) i7-4790 CPU @ 1.3GHz processor and an 8GB memory. We use subset-sum problems (in Section \ref{ProblemContractSample}) as test problems and submit them to the simulation network. A test problem can be defined as: Given a set $S = \{1, 2, 3,{\ldots},n\}$, find a subset of ${S}$ whose sum is ${Z}$ and ${Z=\frac{n(n+1)}{2}-1}$. Thus, the difficulty of a test problem can defined as ${D_n}=\frac{1}{2^n}$. The larger ${n}$ is, the smaller the ${D_n}$ is, and the more difficult the problem is. In our experiments, we take $n=29, 30, 31, 32$ and set the system's benchmark excitation to ${W}$ (in Section \ref{incentivemodel}). For four types of problems with different difficulty (different ${n}$), we provide different reward ${R}$ and ${R{\in}[W,50W]}$. For each combination ${C=\{n, R\}}$, we submit 100 problems separately, and calculate the frequency of successful solving.

\begin{figure}[htb]
  \centering
  \includegraphics[scale=0.55]{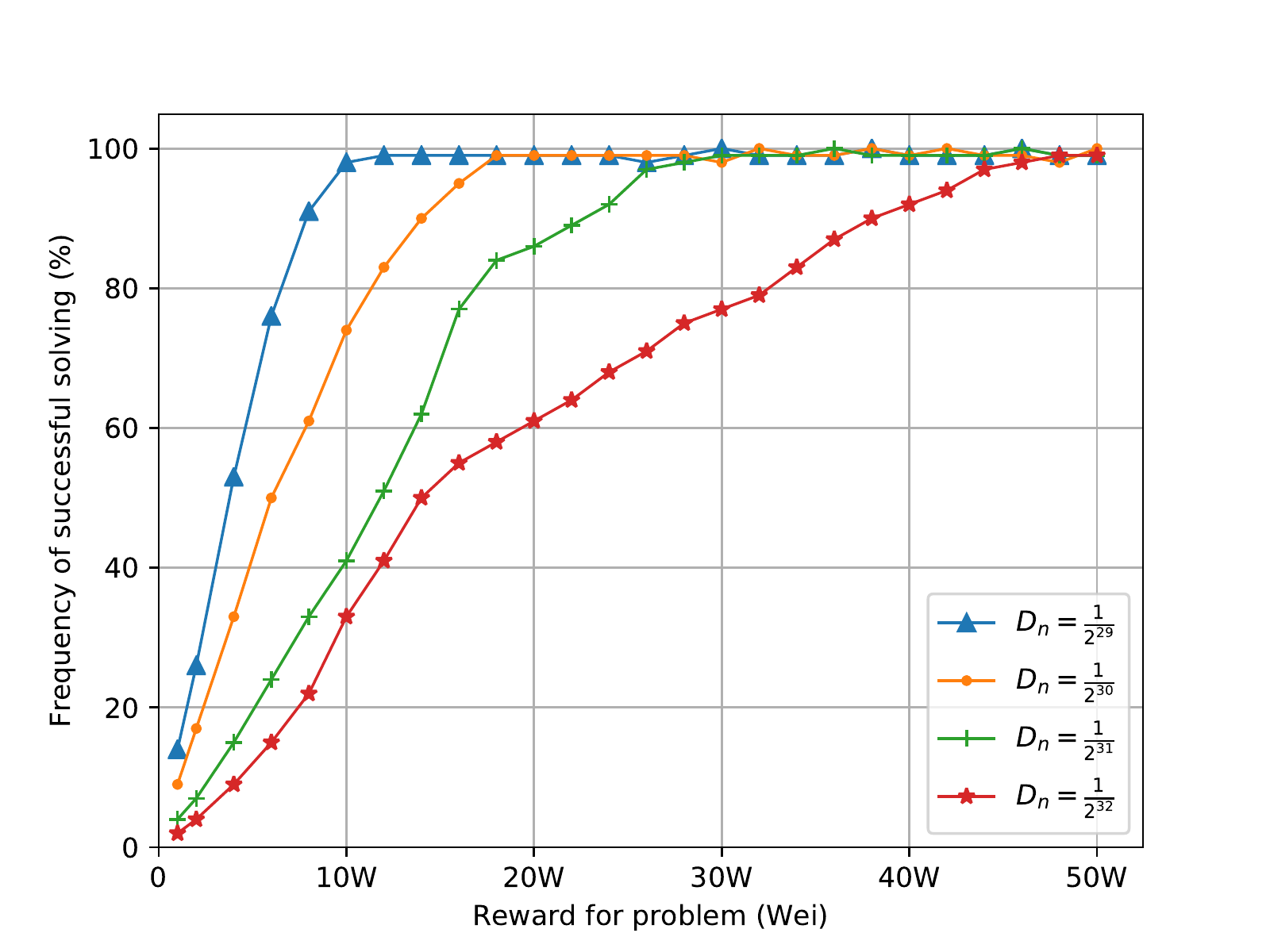}
  \caption{The situation of problem solving}
  \label{problemsolvingpFigure}
\end{figure}

As shown in Figure \ref{problemsolvingpFigure}, for the problem with same difficulty, as the problem’s reward increases, the frequency of successful solving approaches linear growth and eventually approaches 100\%. In other words, as long as the user gives enough reward and the problem is solvable, the problem will be solved by AxeChain. At the same time, the user does not have to worry about providing too much reward, because the extra reward will be returned to the user. In addition, it can be seen that for the same reward, the less difficult the problem (the smaller n), the greater the frequency of successful solving.


\section{Related Work} \label{relatedwork}
Currently, for the energy waste of PoW, the existing studies are mainly divided into two categories. On the one hand, the mainstream alternative is Proof-of-Stake (PoS)\cite{king2012ppcoin, kiayias2017ouroboros}. At present, PoS is still controversial, and there are still many issues that need to be addressed. Nicolas Houy et al. claimed that malicious attackers can easily perform the cost-free simulated attacks on the PoS-based blockchain systems\cite{houy2014will}, while Poelstra claimed that external resource consuming is necessary for blockchain security\cite{poelstra2014distributed}. In addition, PoS will widen the gap between rich and poor. In extreme cases, the rich may control the entire network, leading to centralization.

On the other hand, there are a few studies that translate the hash energy consumption of PoW into meaningful work. For example, Primecoin\cite{king2013primecoin} replaces hash calculation with search of prime sequence and Gapcoin\cite{Gapcoin} searches for large prime gaps. However, the application of prime sequences is very limited. Besides, since the distribution of prime numbers has not been confirmed by humans, the reliability of the difficulty retargeting mechanism of Primecoin cannot be guaranteed. In the study\cite{oliver2017proposal}, Carlos G. Oliver et al. proposed that the hash energy of PoW can be used to solve the NP-complete problems. However, as analyzed in Section \ref{BBGSSecurity}, the proposed consensus mechanism has serious security issues due to \textit{BBGS}. Similar studies also include Hybrid Mining\cite{chatterjee2019hybrid} and Proof-of-Accumulated-Work (PoAW)\cite{amar2019incentive}. Hybrid Mining also uses \textit{BBGS}, so it has the same security issues as the study\cite{oliver2017proposal}. In order to limit the number of problems in the network (prevent attackers from overwhelming the network with many spam problems), Hybrid Mining has greatly increased the cost of submitting problems, which is not friendly to users and reduces the availability of the system. PoAW introduces PoS to form a subnet of miners to manage and verify problems, which will bring security issues of PoS itself. In addition, the reward for solving problems in \cite{oliver2017proposal, chatterjee2019hybrid, amar2019incentive} is completely subjectively determined by users, which can easily lead to unfairness between users and miners. It is worth noting that most of the existing researches focus on solving a certain type of problem (mostly NP problems) and require users to convert the problem into a specific format (such as SAT in Hybrid Mining), so they cannot extend to other application scenarios.

In addition, there are also some related studies in the distributed field. One of the most well-known projects is GIMPS\cite{woltman2007gimps}, which can find huge prime numbers in a distributed manner. Another project is \textit{Folding@Home and Genome@Home}\cite{larson2009folding}, which can solve the hard problem of protein folding that is of great interest to biologists. However, these projects do not provide incentive for contributors and most of the computation is done by volunteers. Therefore, the overall computational power of these projects is far less than that of blockchain. Besides, there is a centralized organization to maintain these project, which runs counter to decentralization of blockchain. 

Compared with the existing researches, AxeChain can use the computing power of blockchain to solve meaningful practical problems while providing strong security guarantees. AxeChain addresses the security issues in the existing researches through \textbf{Axe blockchain architecture} and \textbf{Virtual machine stack sampling proof algorithm}. In addition, AxeChain use Turing-complete high-level language to represent a problem as a smart contract, making the submission of problems more flexible and versatile. The problems submitted to AxeChain are not limited to one type of problem (such as NP problems) and AxeChain can meet a wider range of computing needs. Besides, AxeChain's incentive model can make the blockchain incentive mechanism more practical, which balances the interests of users and miners and provides more value endorsement for cryptocurrency.

\section{CONCLUSION} \label{conclusion}
We present AxeChain, the first Strongly Secure and Fully Decentralized blockchain for solving Easily-Verifiable problems. At its core, AxeChain can translate the hash energy consumption of PoW into useful work while providing strong security guarantees. It is worth noting that the problems that can be submitted to AxeChain will be very diverse (not limited to NP problems), so the computing services provided by AxeChain are more versatile. Additionally, the new incentive model in this paper can make the blockchain incentive mechanism more practical and provide more value endorsement for cryptocurrency.

\bibliographystyle{IEEEtran}
\bibliography{IEEEabrv, sample-base}

\begin{thebibliography}{10}
\providecommand{\url}[1]{#1}
\csname url@samestyle\endcsname
\providecommand{\newblock}{\relax}
\providecommand{\bibinfo}[2]{#2}
\providecommand{\BIBentrySTDinterwordspacing}{\spaceskip=0pt\relax}
\providecommand{\BIBentryALTinterwordstretchfactor}{4}
\providecommand{\BIBentryALTinterwordspacing}{\spaceskip=\fontdimen2\font plus
\BIBentryALTinterwordstretchfactor\fontdimen3\font minus
  \fontdimen4\font\relax}
\providecommand{\BIBforeignlanguage}[2]{{%
\expandafter\ifx\csname l@#1\endcsname\relax
\typeout{** WARNING: IEEEtran.bst: No hyphenation pattern has been}%
\typeout{** loaded for the language `#1'. Using the pattern for}%
\typeout{** the default language instead.}%
\else
\language=\csname l@#1\endcsname
\fi
#2}}
\providecommand{\BIBdecl}{\relax}
\BIBdecl

\bibitem{nakamoto2008bitcoin}
S.~Nakamoto \emph{et~al.}, ``Bitcoin: A peer-to-peer electronic cash system,''
  2008.

\bibitem{natoli2019deconstructing}
C.~Natoli, J.~Yu, V.~Gramoli, and P.~Esteves-Verissimo, ``Deconstructing
  blockchains: A comprehensive survey on consensus, membership and structure,''
  \emph{arXiv preprint arXiv:1908.08316}, 2019.

\bibitem{wood2014ethereum}
G.~Wood \emph{et~al.}, ``Ethereum: A secure decentralised generalised
  transaction ledger,'' \emph{Ethereum project yellow paper}, vol. 151, pp.
  1--32, 2014.

\bibitem{zheng2018blockchain}
Z.~Zheng, S.~Xie, H.-N. Dai, X.~Chen, and H.~Wang, ``Blockchain challenges and
  opportunities: A survey,'' \emph{International Journal of Web and Grid
  Services}, vol.~14, no.~4, pp. 352--375, 2018.

\bibitem{bitcoinenergyconsumption}
{Digiconomist}, ``Bitcoin energy consumption index,''
  \url{https://digiconomist.net/bitcoin-energy-consumption/}, 2017.

\bibitem{electricityconsumption}
{Electricity consumption WIKI}, ``List of countries by electricity
  consumption,''
  \url{https://en.wikipedia.org/wiki/List_of_countries_by_electricity_consumption},
  2014.

\bibitem{king2012ppcoin}
S.~King and S.~Nadal, ``Ppcoin: Peer-to-peer crypto-currency with
  proof-of-stake,'' \emph{self-published paper, August}, vol.~19, 2012.

\bibitem{io2017eos}
D.~Larimer, ``{EOS. IO technical white paper},'' \emph{EOS. IO (accessed 18
  December 2017) https://github.com/EOSIO/Documentation}, 2017.

\bibitem{huobiDENIES}
{OsatoNomayo}, ``Huobi denies involvement in alleged eos voting manipulation,''
  \url{https://bitcoinist.com/huobi-eos-voting-manipulation-denies/}, Accessed
  September 30, 2018.

\bibitem{poelstra2014distributed}
A.~Poelstra \emph{et~al.}, ``Distributed consensus from proof of stake is
  impossible,'' \emph{Self-published Paper}, 2014.

\bibitem{houy2014will}
N.~Houy, ``It will cost you nothing to'kill'a proof-of-stake crypto-currency,''
  \emph{Available at SSRN 2393940}, 2014.

\bibitem{king2013primecoin}
S.~King, ``Primecoin: Cryptocurrency with prime number proof-of-work,''
  \emph{July 7th}, vol.~1, p.~6, 2013.

\bibitem{Gapcoin}
{Sunny King}, ``{Gapcoin},'' \url{https://gapcoin.org/}, Accessed October 21,
  2014.

\bibitem{oliver2017proposal}
C.~G. Oliver, A.~Ricottone, and P.~Philippopoulos, ``Proposal for a fully
  decentralized blockchain and proof-of-work algorithm for solving np-complete
  problems,'' \emph{arXiv preprint arXiv:1708.09419}, 2017.

\bibitem{chatterjee2019hybrid}
K.~Chatterjee, A.~K. Goharshady, and A.~Pourdamghani, ``Hybrid mining:
  exploiting blockchain's computational power for distributed problem
  solving,'' in \emph{Proceedings of the 34th ACM/SIGAPP Symposium on Applied
  Computing}.\hskip 1em plus 0.5em minus 0.4em\relax ACM, 2019, pp. 374--381.

\bibitem{amar2019incentive}
D.~Amar and L.~Zilpa, ``Incentive-based ledger protocols for solving machine
  learning tasks and optimization problems via competitions,'' in
  \emph{Proceedings of the IEEE Conference on Computer Vision and Pattern
  Recognition Workshops}, 2019, pp. 0--0.

\bibitem{MajorityAttack}
{Majority Attack WIKI}, ``Majority attack,''
  \url{https://en.bitcoin.it/wiki/Majority_attack}, 2008.

\bibitem{SHA2}
{National Institute of Standards and Technology(NIST)}, ``Secure hash algorithm
  2,'' \url{https://en.wikipedia.org/wiki/SHA-2}, 2001.

\bibitem{duong2009flickr}
T.~Duong and J.~Rizzo, ``Flickr’s api signature forgery vulnerability,''
  \emph{Tech. Rep.}, 2009.

\bibitem{ethash}
{Vitalik Buterin and Thaddeus Dryja}, ``Ethash[online],''
  \url{https://github.com/ethereum/wiki/wiki/Ethash}, Accessed August 22, 2018.

\bibitem{smith1997application}
M.~J.~S. Smith, \emph{Application-specific integrated circuits}.\hskip 1em plus
  0.5em minus 0.4em\relax Addison-Wesley Reading, MA, 1997, vol.~7.

\bibitem{SHA3}
{National Institute of Standards and Technology(NIST)}, ``Secure hash algorithm
  3,'' \url{https://en.wikipedia.org/wiki/SHA-3}, 2015.

\bibitem{fowler2011fnv}
G.~Fowler, L.~C. Noll, K.-P. Vo, D.~Eastlake, and T.~Hansen, ``The fnv
  non-cryptographic hash algorithm,'' \emph{Ietf-draft}, 2011.

\bibitem{gilad2017algorand}
Y.~Gilad, R.~Hemo, S.~Micali, G.~Vlachos, and N.~Zeldovich, ``Algorand: Scaling
  byzantine agreements for cryptocurrencies,'' in \emph{Proceedings of the 26th
  Symposium on Operating Systems Principles}.\hskip 1em plus 0.5em minus
  0.4em\relax ACM, 2017, pp. 51--68.

\bibitem{kiayias2017ouroboros}
A.~Kiayias, A.~Russell, B.~David, and R.~Oliynykov, ``Ouroboros: A provably
  secure proof-of-stake blockchain protocol,'' in \emph{Annual International
  Cryptology Conference}.\hskip 1em plus 0.5em minus 0.4em\relax Springer,
  2017, pp. 357--388.

\bibitem{buterin2017casper}
V.~Buterin and V.~Griffith, ``Casper the friendly finality gadget,''
  \emph{arXiv preprint arXiv:1710.09437}, 2017.

\bibitem{szabo1994smart}
N.~Szabo, ``Smart contracts,'' \emph{Unpublished manuscript}, 1994.

\bibitem{turing1937computable}
A.~M. Turing, ``On computable numbers, with an application to the
  entscheidungsproblem,'' \emph{Proceedings of the London mathematical
  society}, vol.~2, no.~1, pp. 230--265, 1937.

\bibitem{murty1987some}
K.~G. Murty and S.~N. Kabadi, ``Some np-complete problems in quadratic and
  nonlinear programming,'' \emph{Mathematical programming}, vol.~39, no.~2, pp.
  117--129, 1987.

\bibitem{pardalos1991quadratic}
P.~M. Pardalos and S.~A. Vavasis, ``Quadratic programming with one negative
  eigenvalue is np-hard,'' \emph{Journal of Global Optimization}, vol.~1,
  no.~1, pp. 15--22, 1991.

\bibitem{mordell1969diophantine}
L.~J. Mordell, \emph{Diophantine equations}.\hskip 1em plus 0.5em minus
  0.4em\relax Academic Press, 1969, vol.~30.

\bibitem{sum42}
{ScienceDaily}, ``Sum of three cubes for 42 finally solved -- using real life
  planetary computer,''
  \url{https://www.sciencedaily.com/releases/2019/09/190906134011.htm},
  Accessed September 6, 2019.

\bibitem{Airdrop}
{Airdrop WIKI}, ``Token airdrop,''
  \url{https://en.wikipedia.org/wiki/Airdrop_(cryptocurrency)}, 2017.

\bibitem{harik1999gambler}
G.~Harik, E.~Cant{\'u}-Paz, D.~E. Goldberg, and B.~L. Miller, ``The gambler's
  ruin problem, genetic algorithms, and the sizing of populations,''
  \emph{Evolutionary Computation}, vol.~7, no.~3, pp. 231--253, 1999.

\bibitem{castro1999practical}
M.~Castro, B.~Liskov \emph{et~al.}, ``Practical byzantine fault tolerance,'' in
  \emph{OSDI}, vol.~99, no. 1999, 1999, pp. 173--186.

\bibitem{kotla2007zyzzyva}
R.~Kotla, L.~Alvisi, M.~Dahlin, A.~Clement, and E.~Wong, ``Zyzzyva: speculative
  byzantine fault tolerance,'' in \emph{ACM SIGOPS Operating Systems Review},
  vol.~41, no.~6.\hskip 1em plus 0.5em minus 0.4em\relax ACM, 2007, pp. 45--58.

\bibitem{yin2018hotstuff}
M.~Yin, D.~Malkhi, M.~K. Reiter, G.~G. Gueta, and I.~Abraham, ``Hotstuff: Bft
  consensus in the lens of blockchain,'' \emph{arXiv preprint
  arXiv:1803.05069}, 2018.

\bibitem{chase2018analysis}
B.~Chase and E.~MacBrough, ``Analysis of the xrp ledger consensus protocol,''
  \emph{arXiv preprint arXiv:1802.07242}, 2018.

\bibitem{geth}
{Ethereum}, ``Geth: Official go implementation of the ethereum protocol,''
  \url{https://github.com/ethereum/go-ethereum}, 2015.

\bibitem{woltman2007gimps}
G.~Woltman, S.~Kurowski \emph{et~al.}, ``The great internet mersenne prime
  search,'' \emph{Online],(1997, March 23) available http://www.mersenne.org},
  2004.

\bibitem{larson2009folding}
S.~M. Larson, C.~D. Snow, M.~Shirts, and V.~S. Pande, ``Folding\@ home and
  genome@ home: Using distributed computing to tackle previously intractable
  problems in computational biology,'' \emph{arXiv preprint arXiv:0901.0866},
  2009.

\end{thebibliography}
\appendix
\section{Appendices}

\subsection{Problem Contract Sample}
\subsubsection{\textbf{Diophantine Equation}}
\begin{figure}[htb]
  \centering
  \includegraphics[width=\linewidth]{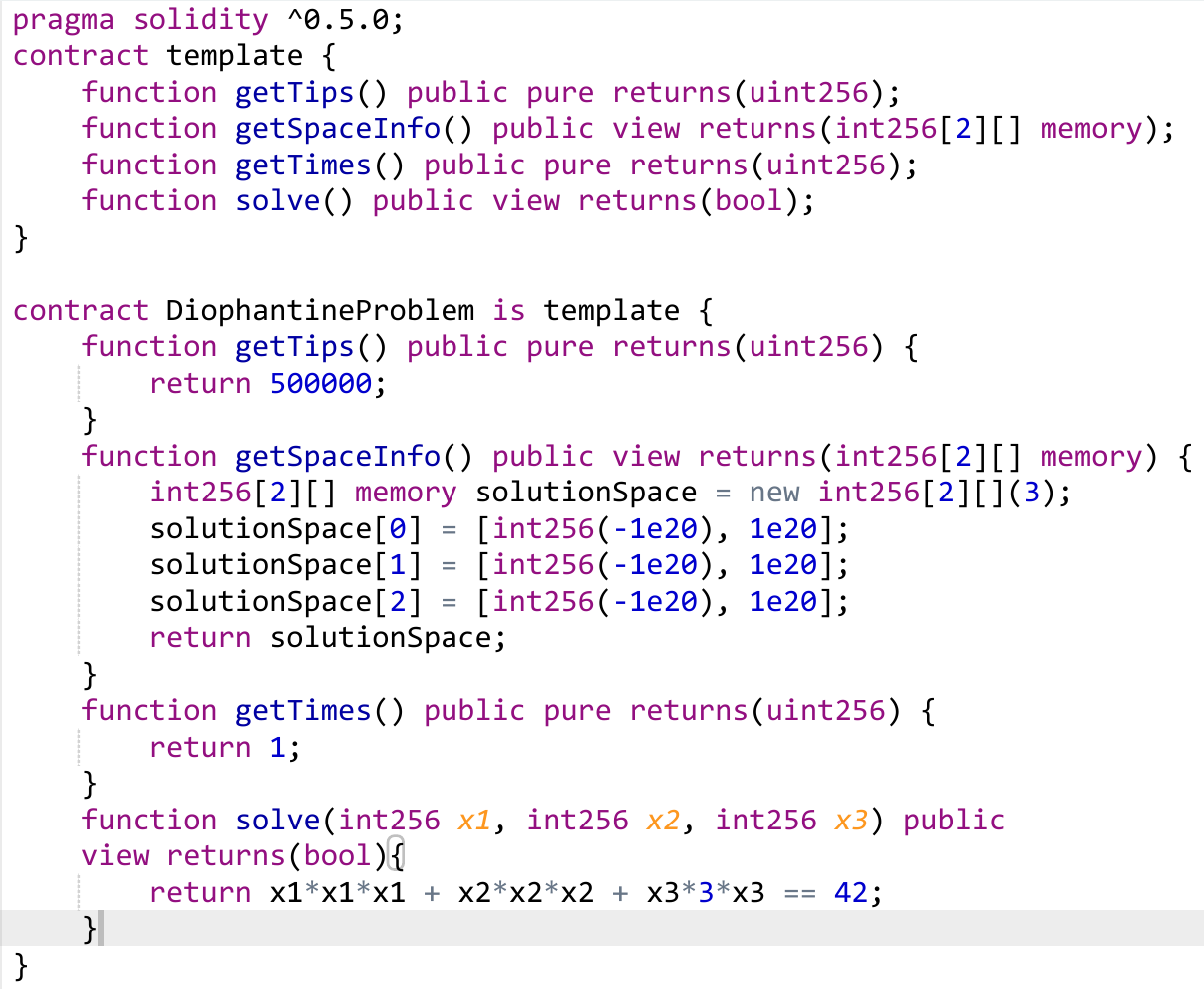}
  \caption{A diophantine problem contract}
  \label{DiophantineProblem}
\end{figure}

The problem contract for finding a solution that Diophantine Equation ${X^3+Y^3+Z^3=42}$ can be shown in Figure \ref{DiophantineProblem}.

\subsubsection{\textbf{Non-linear integer programming problem}}
\begin{figure}[htb]
  \centering
  \includegraphics[width=\linewidth]{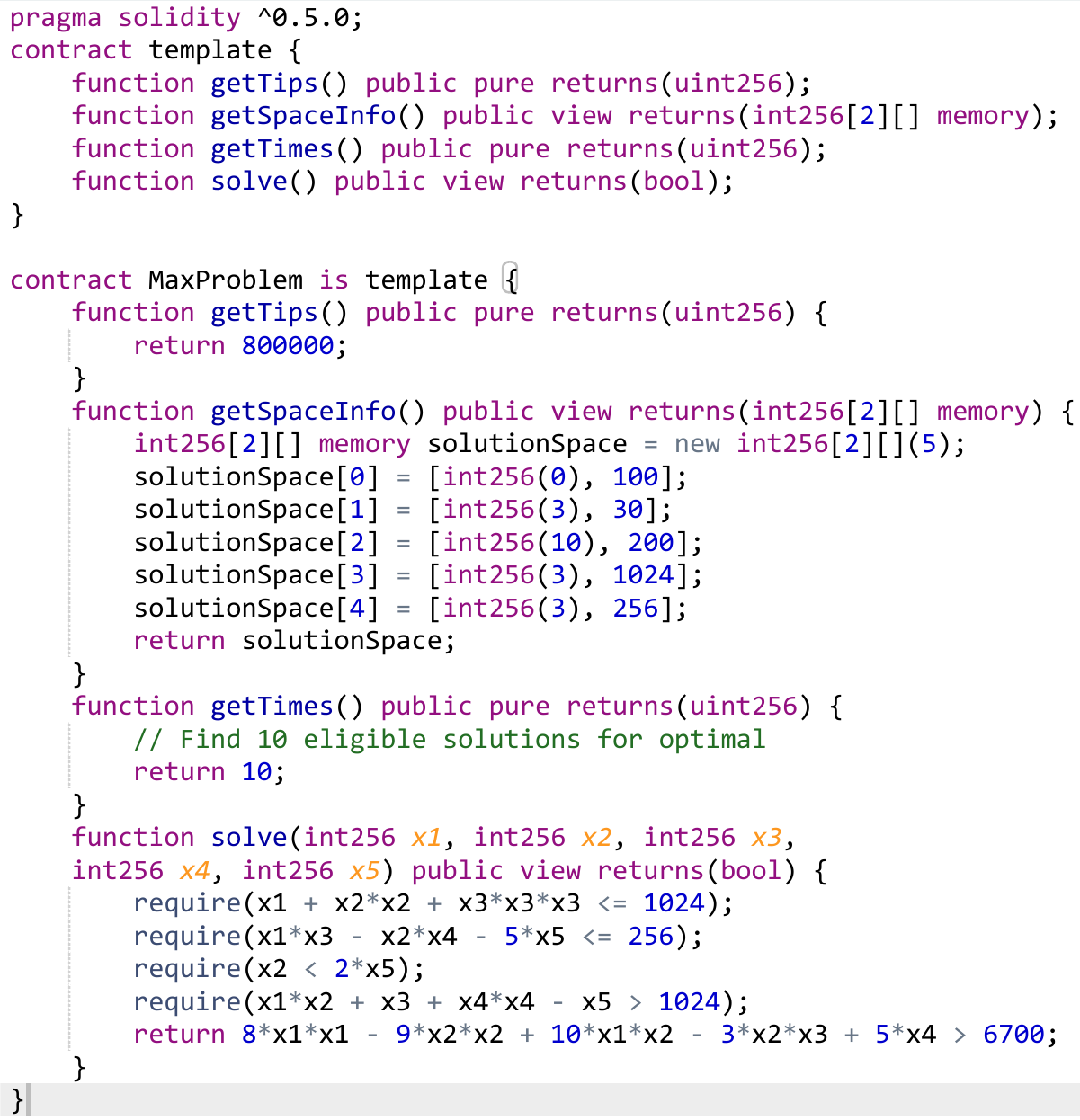}
  \caption{A non-linear integer programming problem contract}
  \label{nonlinear}
\end{figure}

The problem contract can be shown in Figure \ref{nonlinear} and it finds a solution for
\begin{equation}
  \begin{aligned}
    max\,7x^2_1-8x^3_2+9x_1x_2-&3x_1x_2+5x_4-x^2_5 \\
   \left \{
    \begin{aligned}
    x_1+x_2^2+x^3_3\ &{\leq}\ 2014\\
    x_1x_3-x_2x_4-5x_5\ &{\leq}\ 256\\ 
    x_2\ &<\ 2x_5\\
    x_1x_2+x_3+x_4^2-x_5\ &>\ 1024
    \end{aligned}
  \right.
  \end{aligned}
  \label{non-line}
\end{equation}

\subsubsection{\textbf{Token Airdrop}}
\begin{figure}[htb]
  \centering
  \includegraphics[width=\linewidth]{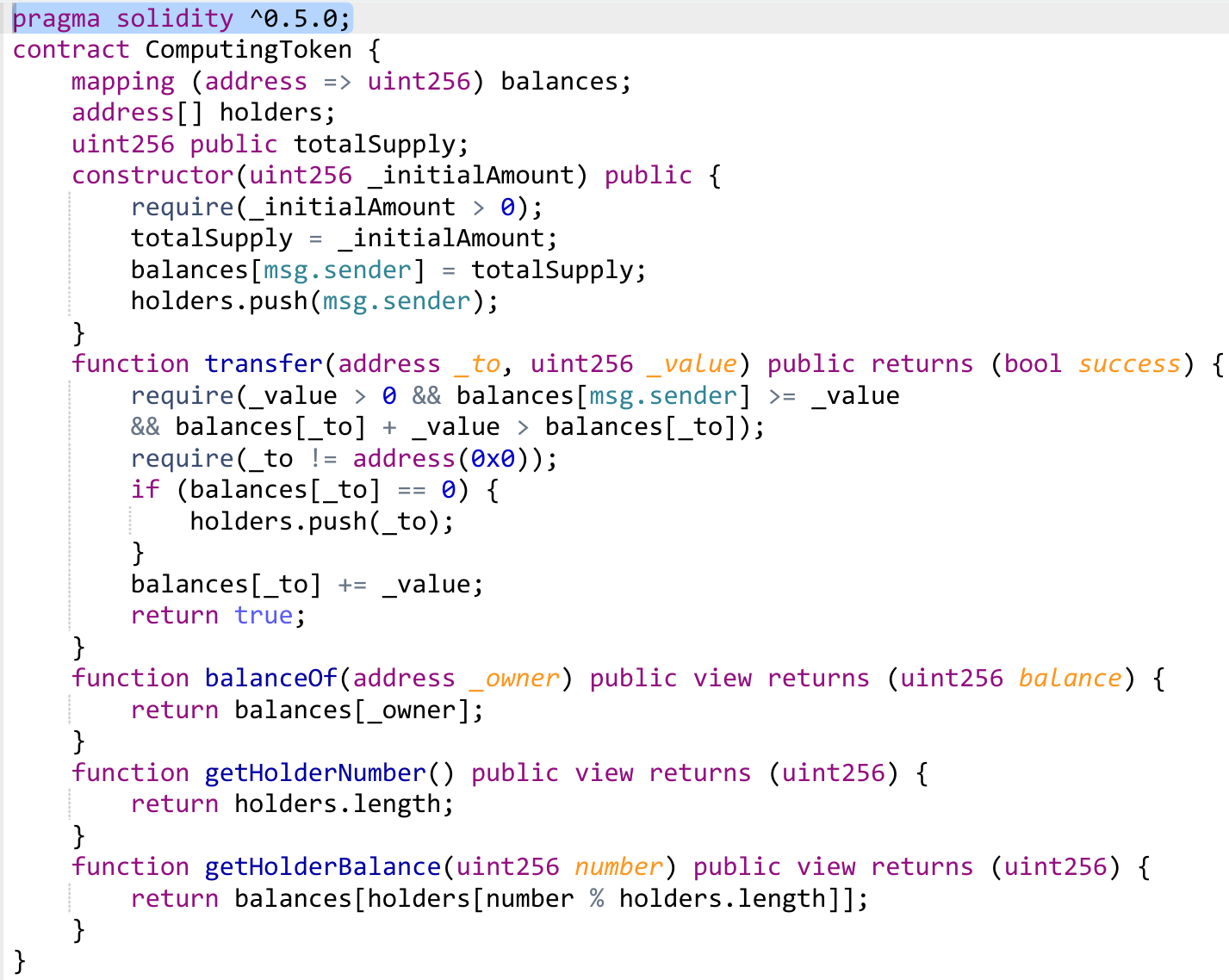}
  \caption{A token issuance contract}
  \label{token}
\end{figure}

\begin{figure}[htb]
  \centering
  \includegraphics[width=\linewidth]{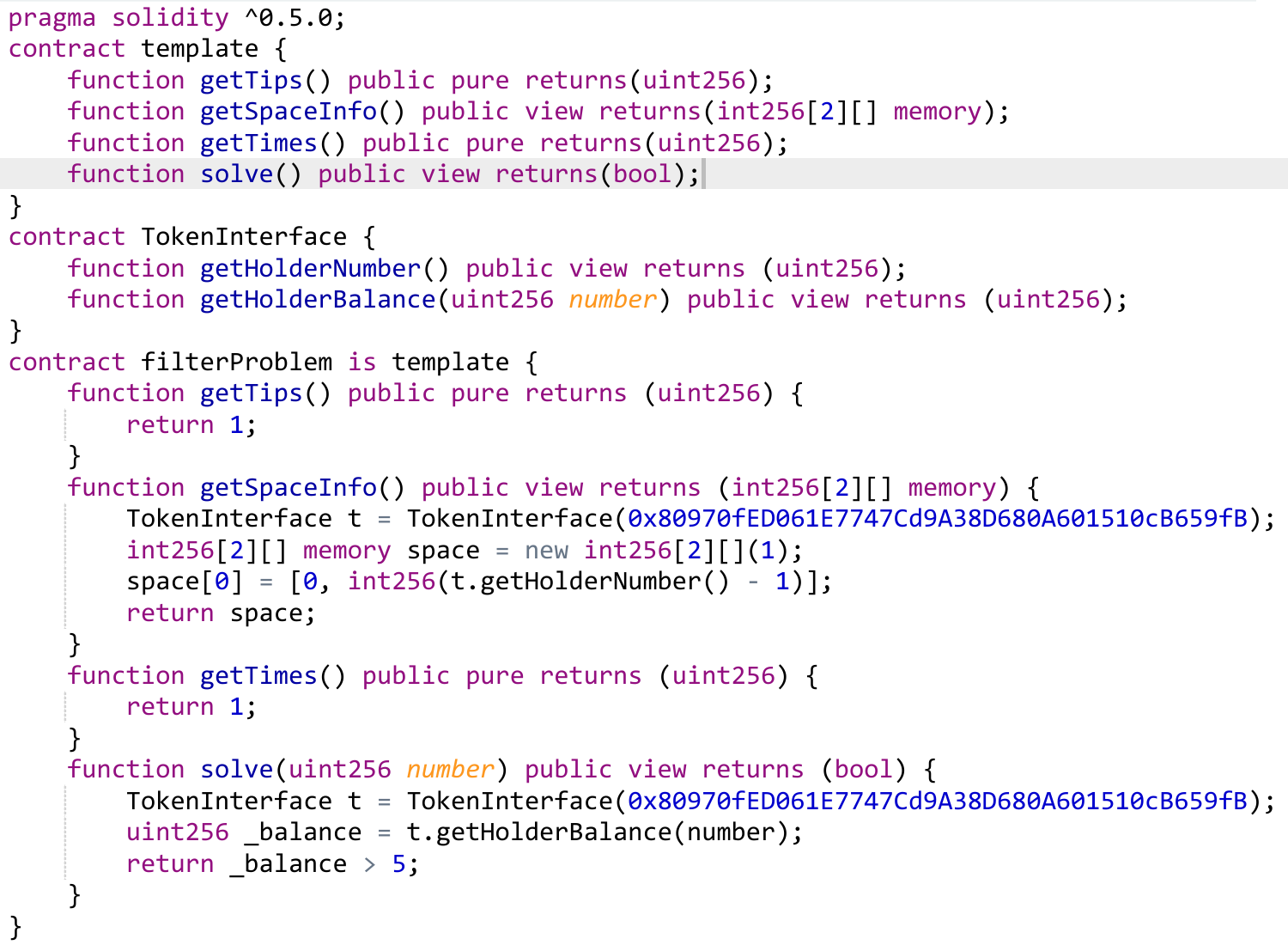}
  \caption{A filtering problem contract}
  \label{filter}
\end{figure}

The blockchain project party performs the initial issuance of the token through the contract \textit{ComputingToken} shown in Figure \ref{token}, and then selects the appropriate token user (active user) through the filtering problem contract \textit{filterProblem} shown in Figure \ref{filter} to perform the secondary issuance of the token. AxeChain helps the project party find the right users quickly and the result is time-sensitive.

\end{document}